\documentclass[sigconf,10pt]{acmart}
\usepackage{amsmath,amsfonts}
\usepackage{mathtools}
\usepackage{algorithmic}
\usepackage{graphicx}
\usepackage{textcomp}
\usepackage{xcolor}
\usepackage{fancyhdr}
\usepackage{chemformula}
\usepackage{graphicx}
\usepackage{setspace}
\usepackage{enumitem}
\usepackage{float}
\usepackage{url}
\usepackage{tabularx}
\usepackage{subfiles}
\usepackage{comment}
\usepackage{wrapfig}
\usepackage{balance}
\usepackage{booktabs}
\usepackage{multirow}
\usepackage{titlesec}
\usepackage{enumitem}
\usepackage{titlecaps}
\usepackage{subcaption}
\usepackage{caption}
\usepackage{xcolor}
\definecolor{highLightChange}{RGB}{34,34,250}

\setcopyright{none}
\renewcommand\footnotetextcopyrightpermission[1]

\titlespacing\section{0pt}{12pt plus 4pt minus 2pt}{0pt plus 2pt minus 2pt}
\titlespacing\subsection{0pt}{12pt plus 4pt minus 2pt}{0pt plus 2pt minus 2pt}

\begin{document}
\title{\emph{Spectral-Loc}: Indoor Localization using Light Spectral Information}
\author{Yanxiang Wang$^{1,2}$,\hspace{3pt} Jiawei Hu$^{1,2}$,\hspace{3pt} Hong Jia$^3$,\hspace{3pt} Wen Hu$^1$,\hspace{3pt} Mahbub Hassan$^1$ \\\hspace{3pt} Ashraf Uddin$^1$,\hspace{3pt} Brano Kusy$^2$,\hspace{3pt} Moustafa Youssef$^4$}

\affiliation{
    \institution{$^1$ University of New South Wales, Australia\\
    $^2$ Data61-CSIRO, Australia \\
    $^3$ University of Cambridge, United of Kingdom\\
    $^4$ Alexandria University, Egypt}
    \country{\tt\small Email:~\{yanxiang.wang, jiawei.hu\}@unsw.edu.au,  hj359@cam.ac.uk \\ \{wen.hu, mahbub.hassan, a.uddin\}@unsw.edu.au\\ brano.kusy@data61.csiro.au, moustafa-youssef@aucegypt.edu}
}
\renewcommand{\authors}{Yanxiang Wang, Jiawei Hu, Hong Jia, Wen Hu, Mahbub Hassan, Ashraf Uddin, Brano Kusy, Moustafa Youssef}

\renewcommand\abstractname{\textsc{ABSTRACT}}
 \begin{teaserfigure}
  \includegraphics[width=\textwidth]{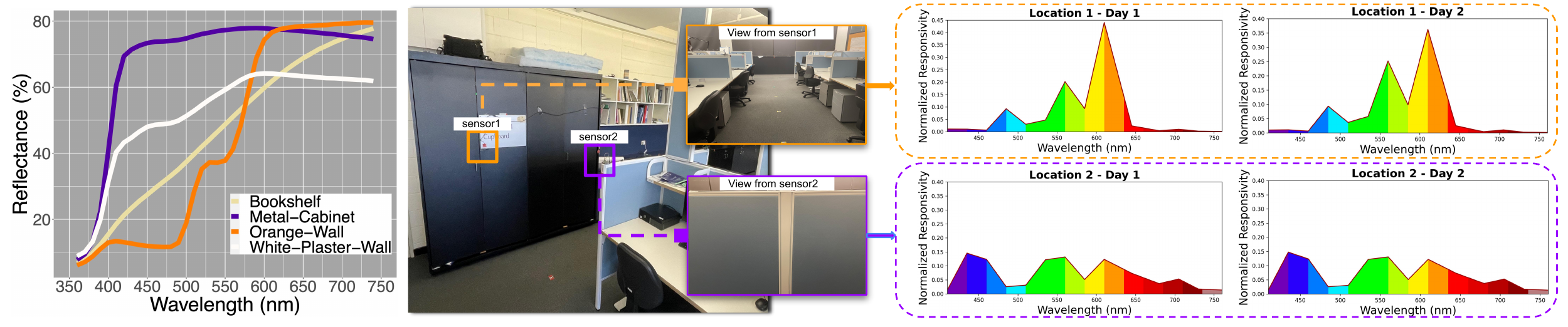}
   \caption{Experimental evidence of the fact that even under the same light source, different indoor locations can observe different light spectral distribution. Different indoor materials have different reflectance curves~\cite{jakubiec2016building} (left), two spectral sensors on the same indoor wall but facing different views (middle), spectral information sensed by the two sensors at two different days (right). }
     \label{fig:teaser}
 \end{teaserfigure}
\begin{abstract}
\noindent
For indoor settings, we investigate the impact of location on the spectral distribution of the received light, i.e., the intensity of light for different wavelengths. Our investigations confirm that even under the same light source, different locations exhibit slightly different spectral distribution due to reflections from their localised environment containing different materials or colours. By exploiting this observation, we propose \textit{Spectral-Loc}, a novel indoor localization system that uses light spectral information to identify the location of the device. With spectral sensors finding their way in latest products and applications, such as white balancing in smartphone photography, \textit{Spectral-Loc} can be readily deployed without requiring any additional hardware or infrastructure. We prototype \textit{Spectral-Loc} using a commercial-off-the-shelf light spectral sensor, AS7265x, which can measure light intensity over 18 different wavelength sub-bands. We benchmark the localisation accuracy of \textit{Spectral-Loc} against the conventional light intensity sensors that provide only a single intensity value. Our evaluations over two different indoor spaces, a meeting room and a large office space, demonstrate that use of light spectral information significantly reduces the localization error for the different percentiles.

\end{abstract}
\maketitle
\pagestyle{plain}
\settopmatter{printfolios=true}
\section{INTRODUCTION}

Indoor localization plays a critical role in many application scenarios, including health-care centers, robot navigation, shopping malls, and smart buildings to name a few. A recent market research  estimates the global indoor localization market to be USD 40.99 billion by 2022~\cite{market}. The significance of this technology has attracted massive research efforts over the past decades, but a ubiquitous solution that would work accurately and reliably in all indoor scenarios is yet to be found~\cite{lymberopoulos2015realistic}. 

Among many radio-based localization options, WiFi-based indoor localization is most extensively studied, mainly due to its ubiquity. However, the ultra-sensitivity of WiFi signals to the surrounding environment often make them unreliable. In addition, a high density of WiFi access points are required to achieve good accuracy, which may not be available in many scenarios. Finally, radio-based applications often face safety concerns in environments like hospitals, mines and military compounds.   

As lighting infrastructure is already there in indoor environments, there is a growing interest to use light for indoor localization. Light is more stable compared to radio and has much higher density of deployments compared to WiFi. Most existing light-based localization solutions, however, rely on modifying the light emitting diode (LED) so that the lighting infrastructure can transmit useful beacons or some spatial information to the receivers for localization~
\cite{kuo2014luxapose,xie2015spinlight,wei2017celli,liu2017smartlight}.
Although modified LED-based techniques can achieve precision and reliable localization, they are limited to LEDs only and incur retrofitting cost as well, which can be a barrier to wide deployments. 

For light-based localization to become ubiquitous and practical, we need solutions that can work with \textit{all types of lights}, LED or otherwise, and \textit{without requiring any modifications or modulations of the light}. However, indoor localization without modulating or modifying the light source is extremely challenging because the receiver can neither identify the light sources, nor it can benefit from any spatial light transmitting patterns anymore. Zhang and Zhang~\cite{zhang2016litell} used the unique resonance frequencies of \textit{fluorescent lamps} to identify them without requiring any modifications, but they do not work for other types of lights such as LED. Since light intensity changes with the distance from the light source, Zhao et al.~\cite{zhao2017navilight} investigated the potential of light intensity to realize indoor localization under arbitrary unmodified lighting infrastructure. However, they discovered that light intensity from a single location has limited performance, hence they proposed a system, called NaviLight~\cite{zhao2017navilight} that uses light intensity values from a series of locations within a trajectory of a walking person for accurate localization. However, trajectory-based localization works only when the user is moving. Accurately fingerprinting single locations under arbitrary unmodified lighting infrastructure remains an open problem.

To address this challenge, we propose to utilize light spectral distribution, i.e., the intensity of light for different wavelengths within the visible spectrum, as a means to fingerprint a given location. The intuition behind this is that different locations are surrounded by different environmental objects, such as walls with different colors, doors with different materials, and so on, which have different light reflection properties. Figure \ref{fig:teaser}(a) shows that walls, bookshelf, and metal cabinets, which are typically found in indoor environments, have quite different spectral reflectance curves. This suggests that, even under the same lighting condition, different locations of the room may exhibit different spectral distribution, which we experimentally verify in Figure \ref{fig:teaser}(b) and (c)). As a result, compared to the scalar light intensity value, use of spectral information may have the potential to achieve more accurate localization under the same lighting conditions.  

We have implemented our proposed spectral information-based localization system, called \textit{Spectral-Loc}, using a commercial off-the-shelf (COTS) light spectral sensor, AS7265x \cite{spectralsensor}, which can measure light intensity over 18 different wavelength frequencies or sub-bands. Our results confirm that light spectral information can be very useful for indoor localization and can significantly outperform solutions that utilize only the light intensity. Given that spectral sensors are becoming a commodity and finding their way in consumer mobile devices~\cite{xiaomi,huawei} for supporting a variety of other applications, \textit{Spectral-Loc} can be readily deployed without requiring any additional hardware or infrastructure.

The contributions of this paper can be summarised as follows:
\begin{itemize}
    \item To the best of our knowledge, we are the first to experiment and analyze the potential of light spectral information for indoor localization with ambient unmodulated light.    
    \item We design machine learning models that can learn location features from spectral data and develop a working prototype for localization called, \textit{Spectral-Loc}, using a COTS spectral sensor. 
    \item We evaluate \textit{Spectral-Loc} by collecting data from two typical indoor environments, a small meeting room and a large open office space, at different days, for varying light conditions, for different number of spectral sensors worn by the user, and for different number of wavelength sub-bands available in each sensor. Our results show that spectral information significantly improves localization accuracy compared to the intensity information under all scenarios. 
\end{itemize}

The rest of the paper is structured as follows. The necessary hypotheses that must hold for spectral information to be used for indoor localization are identified and tested in Section~\ref{sec:hypothesis}. We introduce the background 
model for spectra-based localization in Section~\ref{sec:model} The design of \textit{Spectral-Loc} are discussed in Section~\ref{sec:spectral-loc}, followed by its evaluation in Section~\ref{sec:evaluation}. We discuss the limitations of our implementation and related work in Sections~\ref{sec:discussion} and~\ref{sec:related}, respectively, before concluding the paper in Section~\ref{sec:conclusion}.
\section{HYPOTHESES TESTING}
\label{sec:hypothesis}
In Figure \ref{fig:teaser}, we provided evidence that two locations in the same room under the same lighting condition can observe distinct spectral distributions, which laid the motivation for this paper. In this section, we aim to show that light spectral information not only can be an indicator of location, but it can be a more reliable indicator than the basic intensity information. We achieve this by first identifying the necessary hypotheses that must hold, and then testing them through extensive measurements. We start with a brief background on light intensity and light spectral information.

\begin{figure}
    \centering
    \begin{subfigure}[b]{0.49\linewidth}
    \centering
    \includegraphics[width=\linewidth]{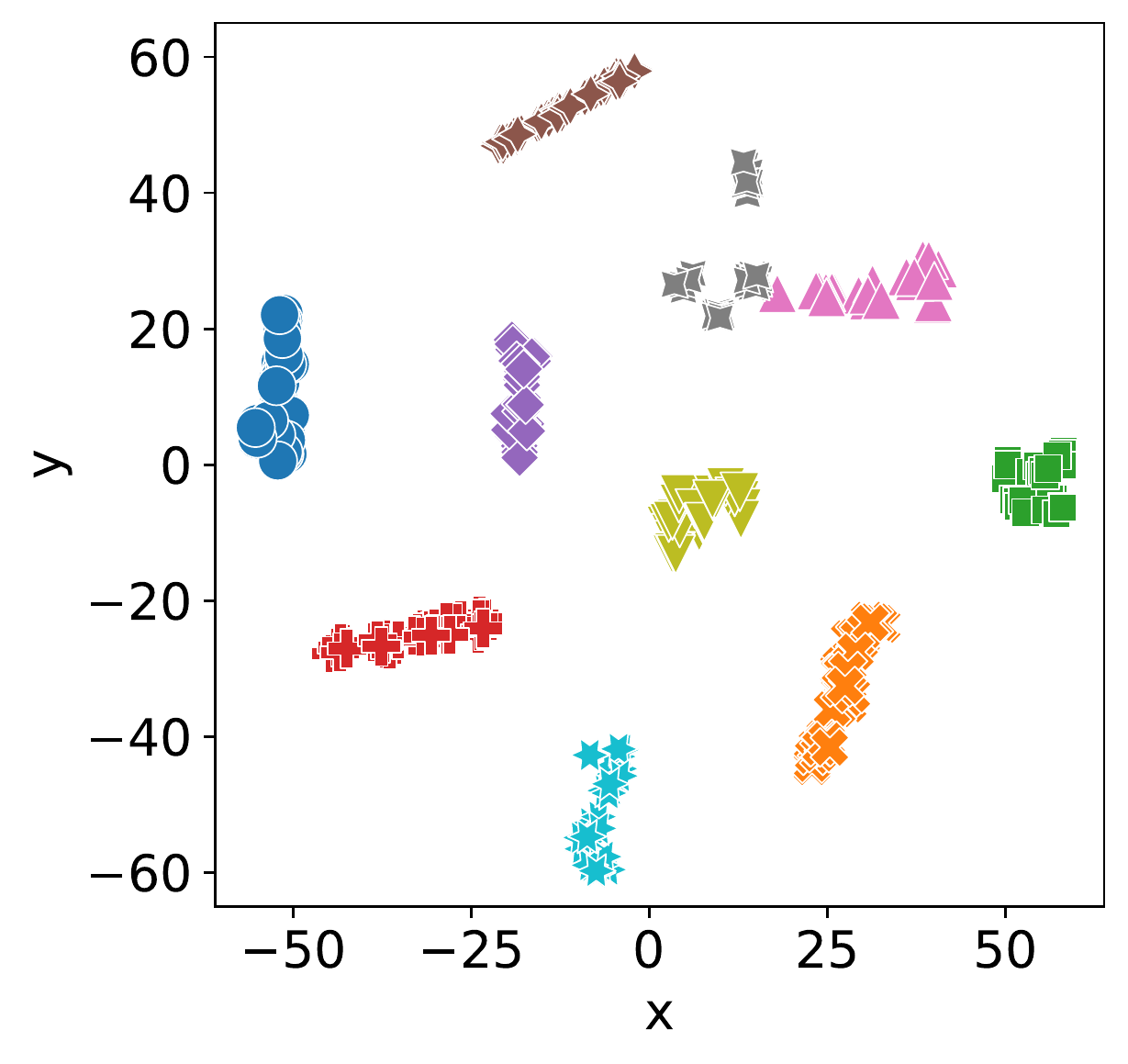}
    \caption{Spectrum}
    \end{subfigure}
    \hfill
    \begin{subfigure}[b]{0.49\linewidth}
    \centering
    \includegraphics[width=\linewidth]{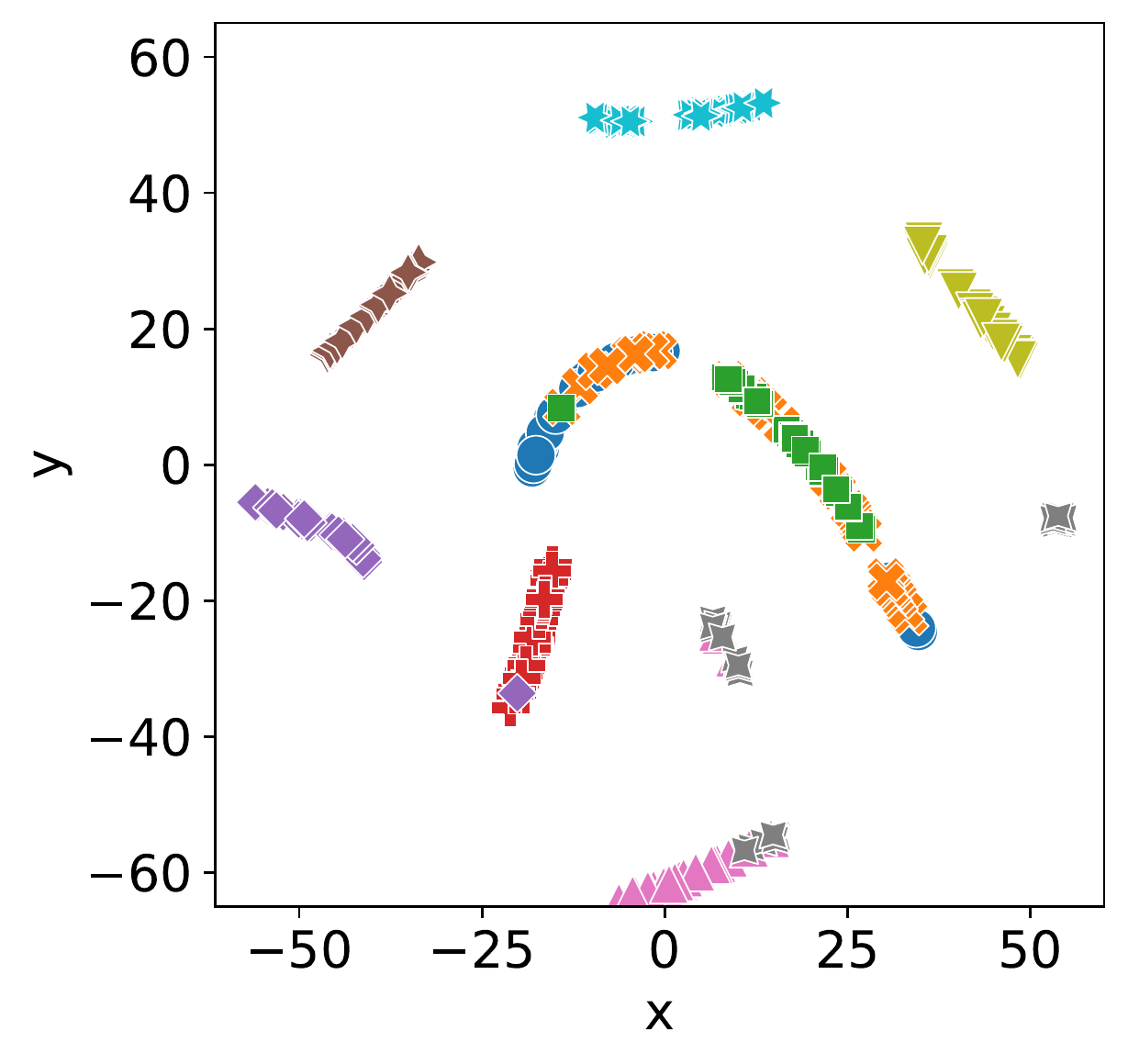}
    \caption{Intensity}
    \end{subfigure}
    \caption{t-SNE visualization of spectral vs. intensity data collected from 10 different locations within a meeting room. The average distance between two clusters is 29.88 and 27.35, respectively, for spectral data and intensity data.}
    \label{fig:tsne}
\vspace{-0.3cm}
\end{figure}

\subsection{Background}

Light intensity is a scalar valued function that returns a single value indicating the received radiant flux per unit area~\cite{wiki}. Light spectrum, on the other hand, is about the color light separation through dispersion systems such as prisms, gratings, or the monochromatic light pattern sequential arranged by wavelength (or frequency). In addition to the overall radiant light power in the area, spectral information also includes the composition of that light source, i.e., the intensity of monochromatic light at each wavelength. In other words, the light spectrum represents the strengths or weights for different wavelengths or frequencies.

While expensive and bulky equipment is required to obtain the full distribution of light over all wavelengths, the recent arrivals of commodity spectral sensors, such as AS7265x that we used in this study and the ones included in some of the latest smartphones~\cite{huawei,xiaomi}, can measure received light over a small set of wavelengths.  These sensors are low-cost low-power devices that have been introduced in the market relatively recently. Some of the applications of these spectral sensors include  product authentication, anti-counterfeiting, portable spectroscopy, adulteration detection, horticultural and specialty lighting, and material analysis~\cite{spectralsensor}.

\subsection{Hypotheses}
We identify three hypotheses that must hold if spectral information is to be useful
(i.e., provide effective features) for indoor localization:
\begin{enumerate}
    \item For typical indoor environments, spectral distribution of received light is location dependent.
    \item Light spectral information is a more reliable indicator of location than basic light intensity information.
    \item For a given environment and lighting condition, spectral distribution of received light at a given location is stable, i.e., preserved over time.
\end{enumerate}

\begin{figure}
    \centering
    \includegraphics[width=6.5cm]{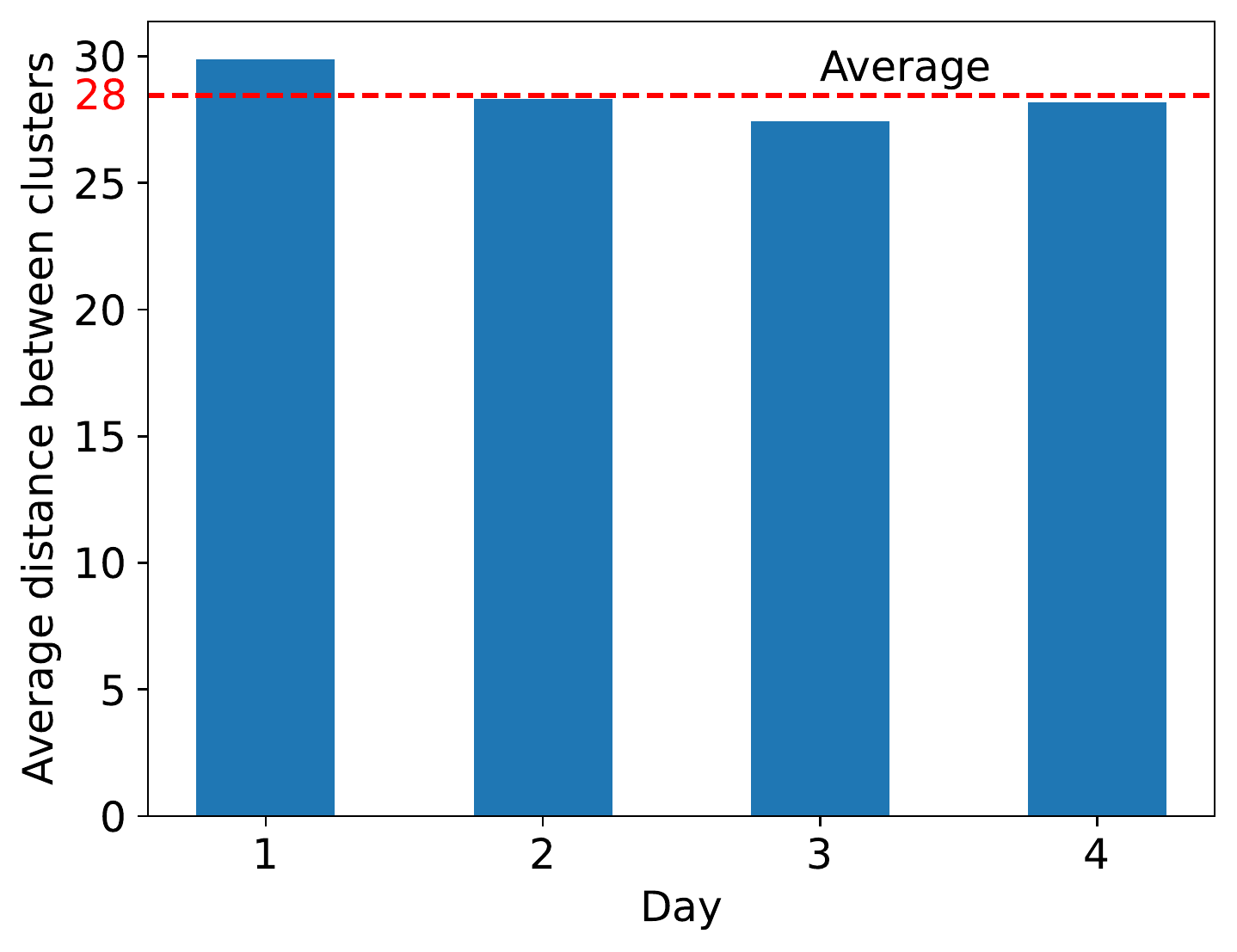}
    \caption{Averaged distance between t-SNE clusters for spectrum information on different days.}
    \label{fig:tsne-days}
\vspace{-0.3cm}
\end{figure}

\subsection{Measurement and testing}
We started by asking a subject to wear an AS7265x at the left wrist and stand for 40 seconds at each of the 10 different locations in a small meeting room. Sensor values from the 18 sub-bands were collected using an Arduino Uno at a rate of 1Hz, which provided 40 18-dimension vectors of spectral data for each location (the details of the room and the sensor implementation will
be discussed in Section~\ref{sec:evaluation}). 

Using t-distributed stochastic neighbour embedding (t-SNE) \cite{van2008visualizing}, we visualize these data, along with the corresponding intensity data that is derived by summing up the 18 spectral values together, in a 2D plot in Figure~\ref{fig:tsne}. The figure shows that spectra data
can indeed differentiate different indoor locations (\textbf{Hypothesis} 1) by representing 
them as clearly separated clusters in Figure~\ref{fig:tsne}(a). Furthermore, the
euclidean distances between any two clusters in Figure~\ref{fig:tsne}(a)
is significantly larger than those in Figure~\ref{fig:tsne}(b), which
indicate that spectral information is a more reliable indicator of location
than basic light intensity information (\textbf{Hypothesis} 2). Finally, we repeat the experiment under the same lighting condition on three other days and observe in Fig.~\ref{fig:tsne-days} that the average t-SNE cluster distances do not change much between days (\textbf{Hypothesis} 3).

 \section{Spectra-based localization model}
 \label{sec:model}

\subsection{Indoor location spectral fingerprint}
Modern buildings are usually equipped with multiple types of light bulbs for indoor lighting. Common lighting includes incandescent light bulbs (ICL), compact fluorescent lamps (CFL), and light-emitting diode (LED). Different indoor environments 
have different layouts, furniture, shadowing, scattering, and light source density and orientation. As a result, the light (intensity 
and spectrum) distributions in an indoor environment is not uniform. Therefore, ambient light intensity and spectrum information may produce stable and discriminative indoor location signatures. We denote the lighting condition in a 2D location $(x, y)$ as:
\begin{equation}
   L(x,y) =  f(\phi),
\end{equation}
where $\phi$ is the ambient light, and $x$ and $y$ are the location coordinates. Earlier work~\cite{zhao2017navilight} has investigated use of basic light intensity value as the indoor location fingerprints:
\begin{equation}
   L_i(x,y) =  f_i(\phi),
\end{equation}
where $f_i(\phi)$ is light intensity of the ambient light in location $(x, y)$, which can
be decomposed as the energy summation of different light spectra:
\begin{equation}
   L_i(x,y) =  f_i(\phi) = \int_{\alpha}^{\beta}(\phi),
   \label{eq:L_i}
\end{equation}
where $\alpha$ and $\beta$ are the starting and finishing light wavelength, respectively. 
Different light intensity sensors respond to different wavelength ranges and have different values for $\alpha$ and $\beta$, which are decided in the sensor manufacturing stage. 

Instead of a single value ($L_i(x,y)$) that represents the total 
energy between a certain wavelength range, i.e., between $\alpha$ and $\beta$
in Eq.~(\ref{eq:L_i}), spectral sensor can return
the energy levels of a number of sub-wavelength ranges (i.e., sub-bands):
\begin{equation}
   L_s(x,y) =  f_s(\phi) = (\phi_i)_{i = 1, 2, ... N},
   \label{eq:L_s}
\end{equation}
where $\phi_i$ is the energy in sub-band $i$, and $N$ is the number
of sub-bands. For example, the COTS
AS7265x spectral sensor, used in the \emph{Spectral-Loc} prototype in
this paper, supports $N = 18$ sub-bands, ranging from ultraviolet (UV) light (i.e., 410 nm) to infrared (IR) light (i.e., 940 nm), with approximately 30 nm wavelength range in each sub-band. In essence, spectral sensors can measure finer grain ambient light information (e.g., the different colors of reflective objects such as walls) compared to the basic light intensity sensors.

By following the localization methodology proposed by Sen et. al. in~\cite{sen2012spinloc}, we divide the indoor
space in the granularity of 1m $\times$ 1m boxes (i.e., spots), 
and the spectra-based localization problem becomes
\begin{equation}
    P(x_{i},y_{i}) = F(C|f_s(\phi)),
\end{equation}
where $P(x_{i},y_{i})$ is the predicted location coordinate, and $C|f_s(\phi)$ is the real time (current) spectral fingerprint
observation. 
The prediction function $F$ is produced automatically via machine learning 
training process based on historical spectral fingerprints,
which will be discussed in details later in Section~\ref{sec:spectral-loc}.

\subsection{Light sources}
\label{subsec:lightsources}
\begin{figure}[htp!]
    \centering
    \includegraphics[width=8cm]{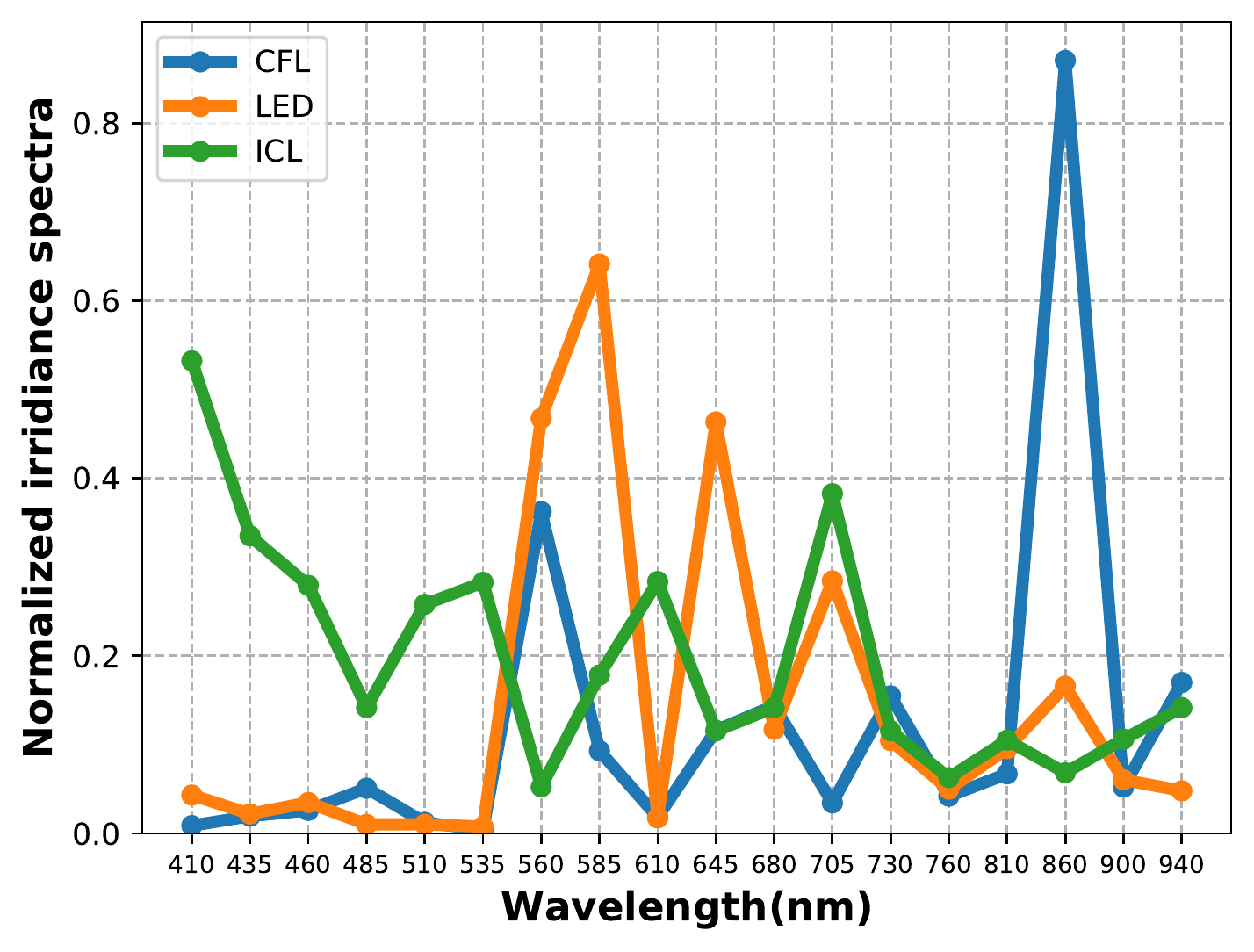}
    \caption{Different spectrum distributions from three common types of light bulbs.}
    \label{fig:difflight}
\vspace{-0.5cm}
\end{figure}

As discussed in the previous section, there are three main types of light bulbs in the market: ICL, CFL, and LED. The ICL light bulbs produces lights by heating a wire filament, and the light color is closer to yellow. Different to ICL light bulbs, CFL light bulbs generate invisible UV lights by charging a mercury vapour. Then, the UV lights hit the fluorescent coating in the tube to produce visible lights. As for LED, it is semiconductor-based light bulbs that release lights when the electrons are combined with the electron holes. The energy states of electrons and electron holes in different semiconductor materials are different. The more the released energy, the shorter the wavelength of the emitted lights. Because of their different 
mechanisms of light  generation, the lights generated by different types of light bulbs have distinct patterns of spectrum distributions, which is shown in Fig.~\ref{fig:difflight}. Since light bulbs are relatively fixed, these
distributions can be exploited by (both intensity and spectra-based) localization
algorithms to produce unique fingerprints (i.e., $L_i(x,y)$ and $L_s(x,y)$).

\subsection{Light reflection}
\label{subsec:lightrefection}
Light reflection is a physical phenomenon that occurs when light enters from one kind of medium into another, and its propagation direction changes at the interface of the two media before returning to its original medium. The reflectance of materials indicates their effectiveness in reflecting radiant energy. Normally, the reflectance is related to the light's incident angle, polarization, and wavelength (frequency). Specifically, the spectral reflection curve describes the material's reflection characteristics for different wavelengths, which can be formulated~\cite{ISO9288} as:
\begin{equation}
    R_{\lambda }=\frac{L^{r}_{\lambda}}{L^{i}_{\lambda}},
\end{equation}
where $R_{\lambda }$ is the reflectance in wavelength $\lambda$,  $L^{r}$ is the spectral radiance of reflected light, and $L^{i}$ is of incident light. 

For different materials and different colors, the reflectance is different. For example, the red interior wall has higher reflectance values for the red light range (622  $\sim $ 780 nm) and lower reflectance ratios for the rest. This imbalance indicates different objects will selectively absorb specific wavelengths of lights, resulting in differences in the wavelengths of the reflected light, which is the reason that the human eyes and cameras can perceive different colors of objects. In the database~\cite{jakubiec2016building}, there are 1,294 types of opaque materials' reflectance measured with spectrally-specific spectrophotmeter sensing devices. The left subplot in Fig.~\ref{fig:teaser} at the start
of this paper show the reflectance curves of four classes of materials (i.e., bookshelf, metal cabinet and two different colors of walls), which are commonly found in different indoor environments, from the database. The figure shows that the reflectance characteristics of them are significantly different to each other, which will produce non-identical reflections even with the same light sources.

To verify our observation that reflections will change the spectrum distributions of lights, we deployed a number of spectral sensors (AS7265x) in a room with the same light sources. The sensors' locations are shown in the middle subplot in Fig.~\ref{fig:teaser}. %
We collected light spectrum measurements from the sensors for different days. The right subplot in Fig.~\ref{fig:teaser} shows that the spectrum distributions in
different locations are indeed unique in different locations, as a light reflection result of the different surrounding objects (e.g., furniture in different colors). Furthermore, the
spectrum distribution is consistent in different days, which 
make them a good feature candidate for fingerprint-based indoor localization
algorithms.

\subsection{The spectrum readings}
\label{subsec:spectramreadings}
\begin{figure}[htp!]
    \centering
    \includegraphics[width=6.5cm]{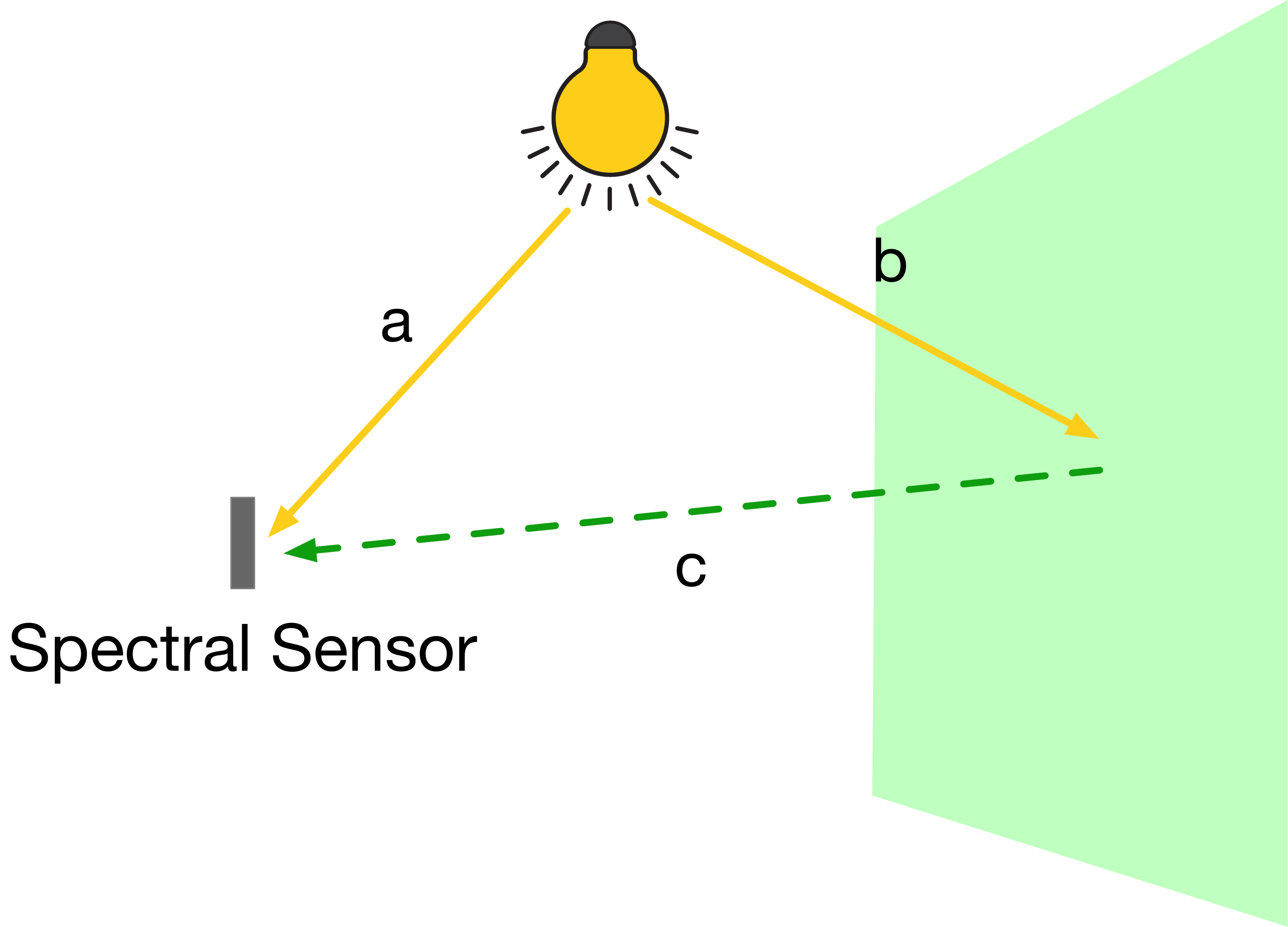}
    \caption{Illustration of light propagation.}
    \label{fig:lot+relection}
    \vspace{-0.5cm}
\end{figure}

The ambient lights measured by spectrum sensors consist of two main components:
direct lights from light sources (e.g., light bulbs) discussed in Section~\ref{subsec:lightsources} earlier,
and reflective lights from surrounding objects discussed in Section~\ref{subsec:lightrefection}. We denote the lights from the light sources 
directly and the reflective lights as $L_{LOS}$ and $L_{RE}$, respectively. 
Both $L_{LOS}$ and $L_{RE}$ follow the inverse-square law for visible light propagation. 
For the time being, let us assume there is one light source and one reflective object in the environment (see Fig.~\ref{fig:lot+relection}). Then, we can model the spectrum sensor's measurements 
in wavelength $\lambda$ as a function of the light source  and
reflective objects as:
\begin{equation}
\begin{aligned}
      L_{\lambda} &= L_{LOS}^{\lambda}+ L_{RE}^{\lambda}\\
   &=\frac{L_b^{\lambda}}{a^{2}}+\frac{1}{c^2}\cdot R_{\lambda}(\frac{L_b^{\lambda}}{b^{2}}),
\end{aligned}
\end{equation}
where $L_b^{\lambda}$ is the energy of the light source in wavelength
$\lambda$ in one of the directions, $a$ is the direct distance between the light source and
the sensor, $b$ is the distance between the light source and the reflective object, and $c$
is the distance between the reflective object and the sensor.

Based on this simplified model, we can infer that the sensor readings from a single light source is not uniformly distributed in different locations because the distances ($a$, $b$, and $c$)
are different. %
As a result, the spectrum measurements are not uniformly distributed. We note that more 
light sources and reflective objects  will only make the measurements more 
uneven. Compared to the intensity measurements, the spectrum measurements
are also affected by the reflectance function $R_{\lambda}$ of surrounding objects, which
is a function of the colors and materials as discussed earlier in Section~\ref{subsec:lightrefection}. This makes it more unique in different 
locations; therefore, more suitable as location fingerprints. 

\section{\textit{Spectral-Loc}: Localization based on spectrum information}
\label{sec:spectral-loc}
\begin{figure}[htp]
    \centering
    \includegraphics[width=7.5cm]{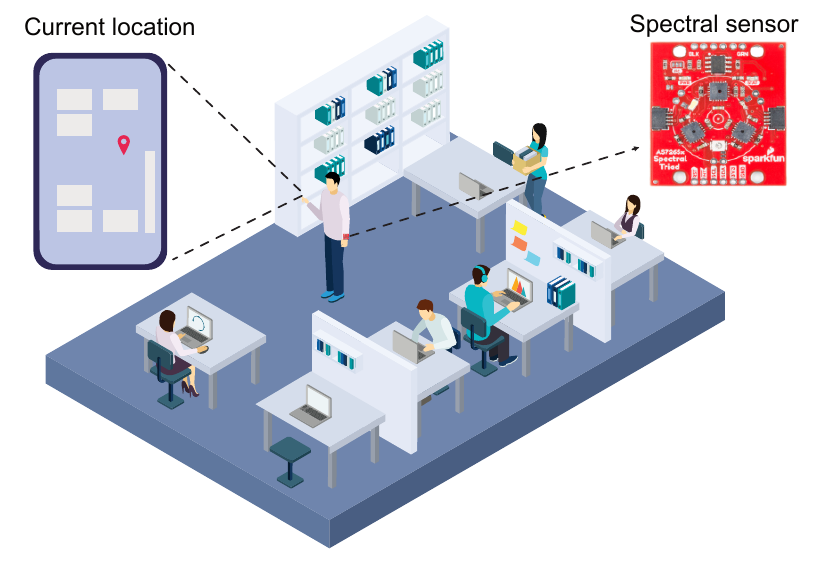}
    \caption{Illustration of indoor localization using spectral information captured by spectral sensors.}
    \label{fig:sketch}
\end{figure}

\begin{figure}[htp!]
    \centering
    \includegraphics[width=\columnwidth]{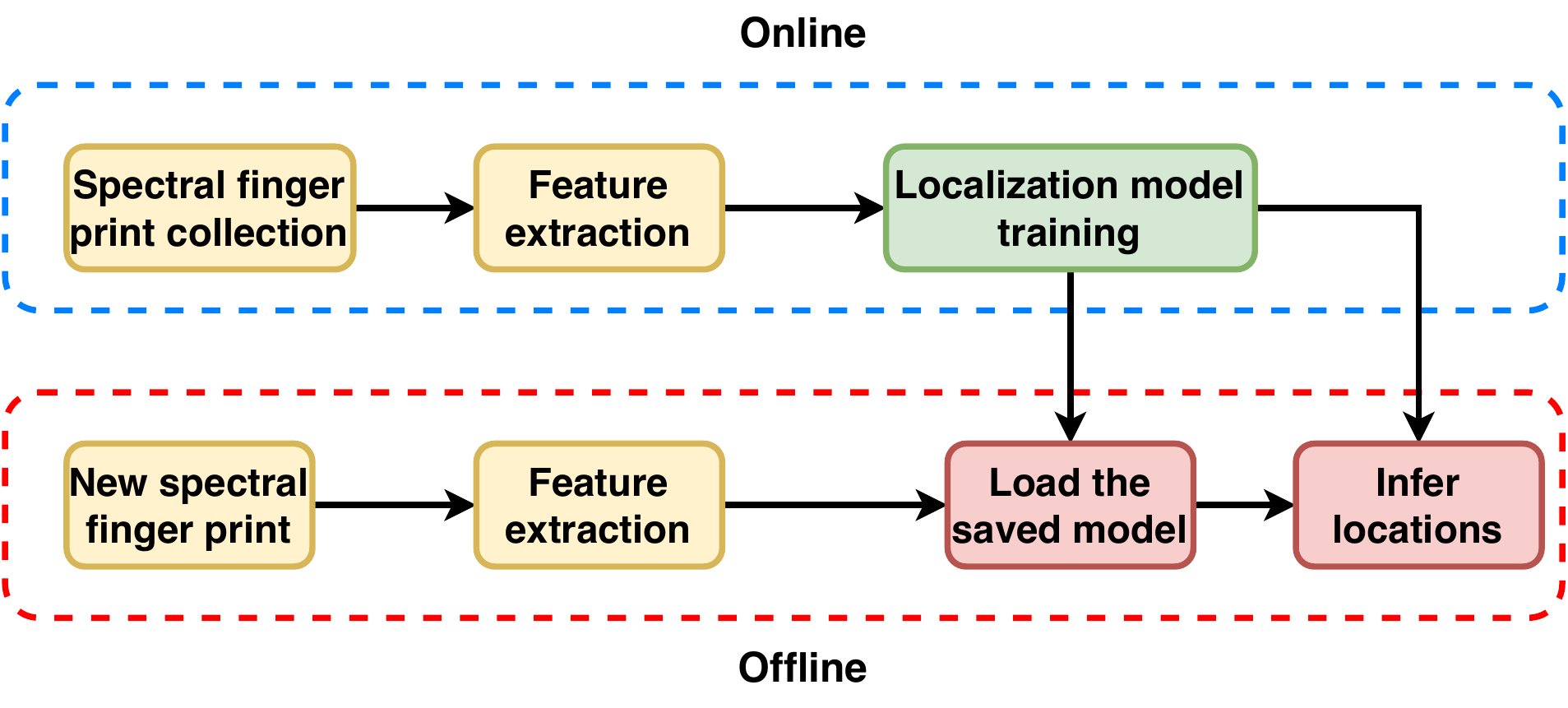}
    \caption{System overview}
    \label{fig:architecture}
\end{figure}

\emph{Spectral-Loc} estimates a user's current location in a room based
on the measurements from spectrum sensors worn by her/him (see Figure~\ref{fig:sketch}). It has two phases: online training and offline location inferencing (see Fig. \ref{fig:architecture}). In the online training phase, the spectral fingerprints are collected at different location points (``spots'') in a room. Then, the data, i.e., the fingerprints ($f_s(\phi)$) with their labels ($x$, $y$), are transferred to a server, which will train the localization module. In the offline testing phase, a new spectral fingerprint  ($f_s(\phi)$) is measured by the spectrum sensor worn by a user in real time and sent to the server, which will predict the user's current location ($x$, $y$) based on 
the fingerprint and the trained localization model from previous phase.

\emph{Normalization.} 
Before we input the spectral sensor measurements (fingerprints, or $f_s(\phi)$) to the 
the localization model, we perform the normalization on a measurement ($\phi_{i}$) by:
\begin{equation}
    \phi^{*}_{i}=\frac{\phi_{i}-\phi_{min}}{\phi_{max}-\phi_{min}}
    \label{e:norm}
\end{equation}
where $\phi_{max}$ and $\phi_{min}$ are the maximum and minimum values in the fingerprint measurements respectively, and $\phi^{*}_{i}$ is the normalized measurement. As will be shown later in Section~\ref{s:fail}, the proposed normalization of spectral readings improves \emph{Spectral-Loc}'s robustness against `unseen' lighting conditions.

\begin{figure}[htp]
    \centering
    \includegraphics[width=\columnwidth]{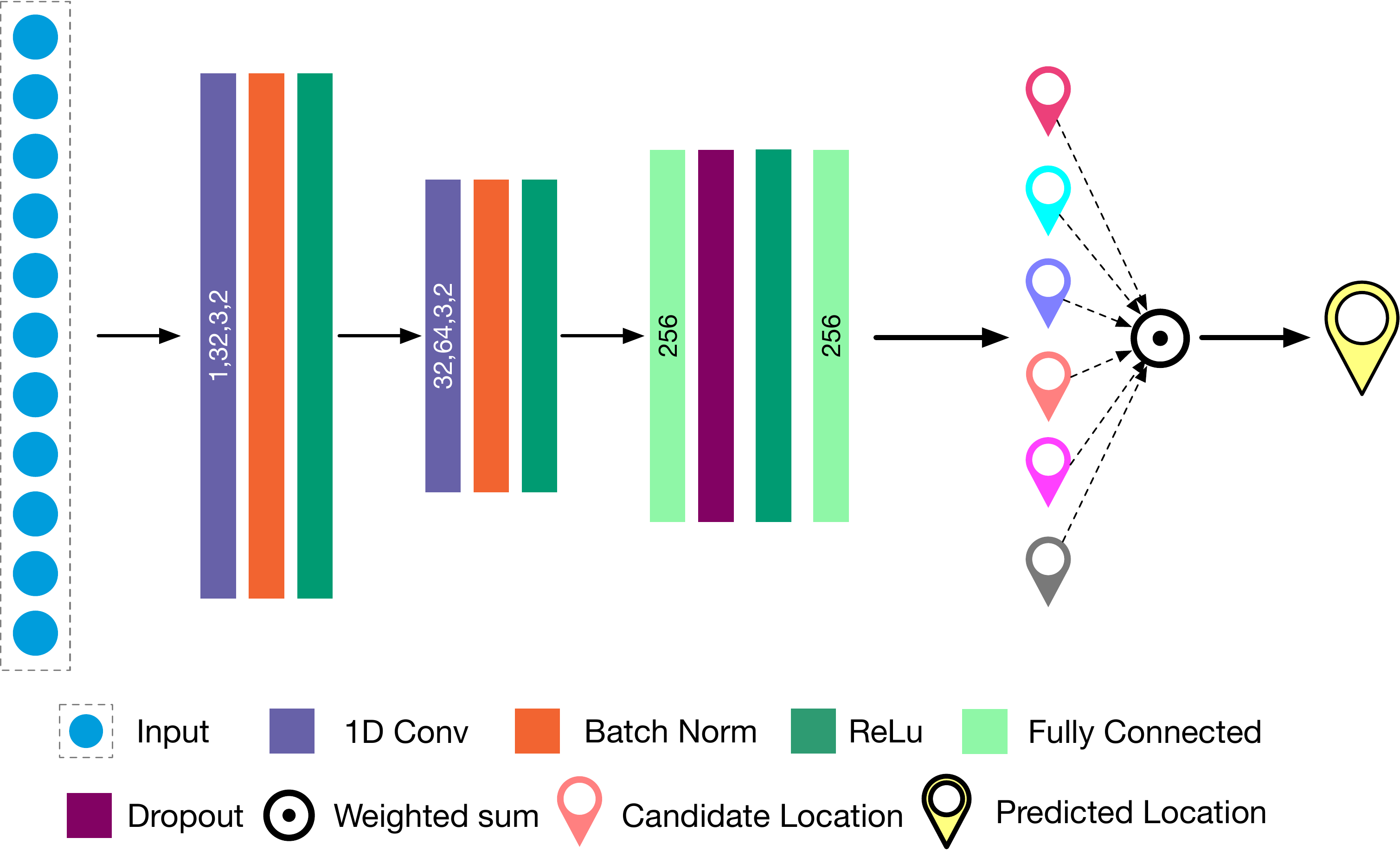}
    \caption{\emph{Spectrum-Loc} deep neural network model.}
    \label{fig:network}
\end{figure}

\emph{Localization neutral network model.}  Our model is based on Convolutional Neural Network (CNN) with an attention mechanism to consider the contributions from all possible locations. Fig.~\ref{fig:network} shows the detailed network model architecture,
key parameters and their values.
The input is the normalized measurements ($\phi^{*}$) from the spectral sensors. If we have $M$ sensors while each sensor has $N$ sub-bands of wavelengths, the input dimension will be $M \times N$. Next, we stack two 1D convolutional layers to extract the
location informative spectrum features. The input channels of the layers changes from 1 to 32 and 32 to 64 respectively, with kernel size of 3 and stride step of 2. After every convolutional layer, the Relu activation~\cite{agarap2018deep} and batch normalization layer are employed. The Relu activation layer can improve the network's non-linearity, which represents the physical environments better, while the batch normalization~\cite{ioffe2015batch} makes the network training faster and more stable. Besides, to prevent over-fitting, we add a dropout layer between the two fully connected layers. The last fully connected layer generates the weights for all possible locations, which, in the next step, are used to calculate the predicted location coordinates ($x$, $y$). In our model, the batch size is 32, the learning rate is 1e-5, and the optimizer is Adam~\cite{kingma2014adam}.

\section{EVALUATION}
\label{sec:evaluation}

\subsection{Goals, Metrics and Methodology}
\label{subsubsec:methoology}
Our goal is to show whether \textit{Spectral-Loc} can improve the performance of indoor localization compared to conventional light intensity-based localization.  The metrics include localization error statistics such as median, 75th percentile
and 90th percentile errors. The evaluation methodology is described below.

\begin{figure}
    \centering
    \includegraphics[width=\columnwidth]{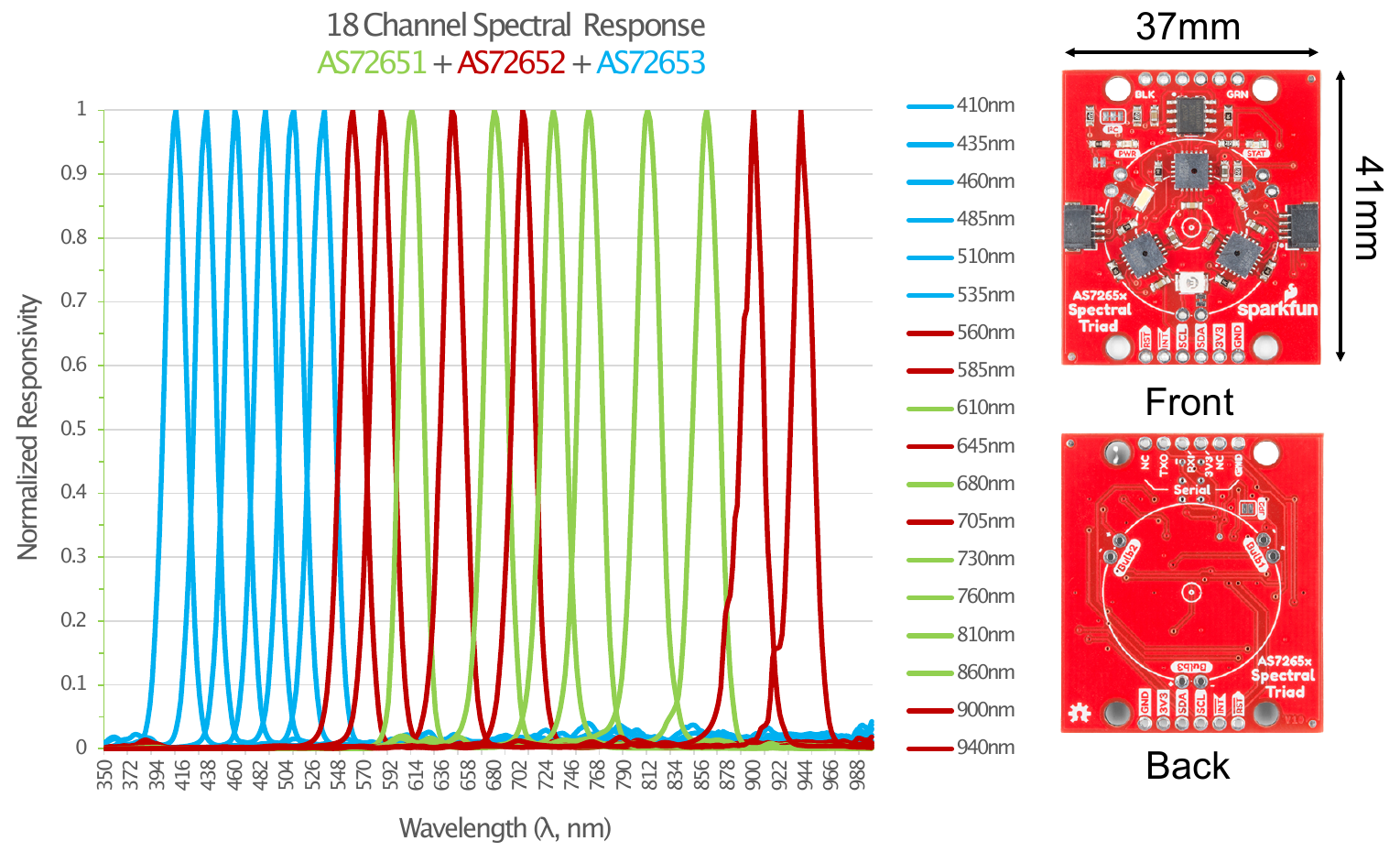}
    \caption{Spectral responsiveness and brief illustrations of AS7265x, which is reproduced from ~\cite{spectralsensor,as7265x}.}
    
    \label{fig:spectral_sensor}
    \vspace{-0.5cm}
\end{figure}

\textbf{Hardware prototype.} 
To measure light over the entire visible spectrum, we use AS7265x~\cite{spectralsensor}
as the hardware prototype of \textit{Spectral-Loc}, which integrates three individual spectral sensors, AS72651-AS72653, for covering the three primary colors, GREEN, RED, and BLUE, respectively. Each individual sensor measures light from 6 different wavelengths within its primary color, thus measuring a total of 18 wavelengths or sub-bands within the spectrum ranging from 410 nm (UV) to 940 nm (IR). In each sub-band, the sensors can measure lights with precision down to 28.6 $nW/cm^{2}$ and accuracy of $\pm $12\%. The sensors' normalized responsivity for different wavelengths is shown in Fig. \ref{fig:spectral_sensor}. The whole sensor board has a compact size of 41mm $\times$ 37mm with 5mm thickness and 9.98 gram weight, which allows it to be easily carried by a person for experiments. We use an Arduino UNO to sample the spectral values from AS7265x at a rate of 1Hz and transmit them to a Raspberry Pi through USB cables. Each spectral sample, therefore, contains a vector of 18 elements representing light from the 18 sub-bands (wavelength range).

\begin{figure}[htp!]
    \centering
    \includegraphics[width=8cm]{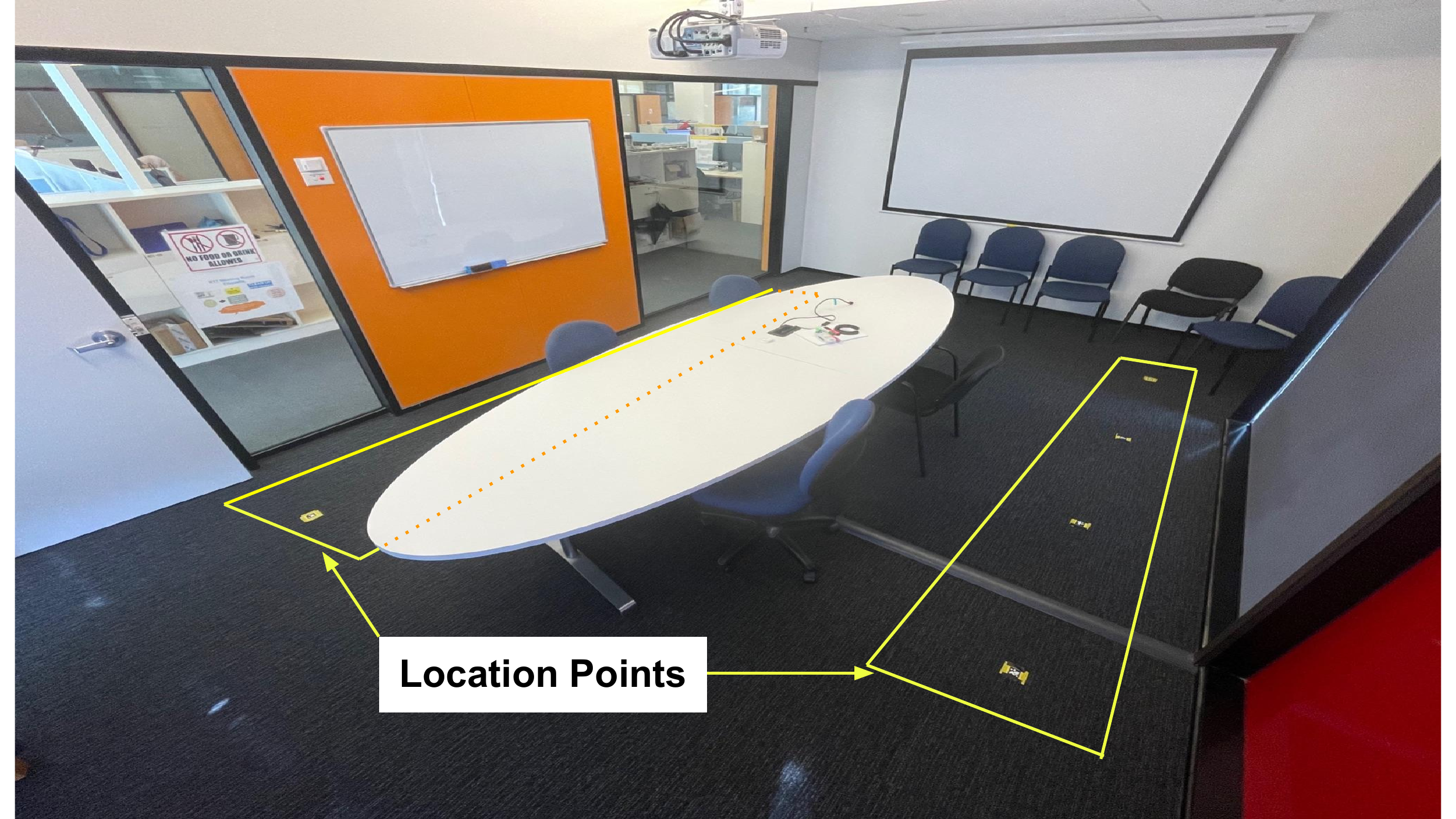}
    \caption{The layout of the meeting room.}
    \label{fig:401K}
    \vspace{-0.5cm}
\end{figure}

\textbf{Experiment environment.}
To evaluate the robustness of \textit{Spectral-Loc}, we collected data from two different typical indoor environments: a small meeting room and a large open office space, which have different layouts, lighting, wall colors, and furniture. %
The \textbf{meeting room} is a 7.36m $\times$ 3.91m rectangular room with clear glass walls to the adjacent open office area. 
The room has two colored walls and a black carpet on the floor.  In terms of lighting, the room is lit by two rows of Crompton T8 36W 840 fluorescent tube lights on the ceiling to produce diffused lights to the room. 
Furthermore, the light in the adjacent open office area affects the lighting conditions in the meeting room via the glass walls.
Finally, we used a 1.4m high 420lm floor lamp with a switchable colour temperature between 2,000K (White) and 4,500K (Cool White), to provide
further lighting control in the meeting room. Using April Tags \cite{olson2011tags} with identifying bar code, we marked 10 locations on the floor spreading in the room evenly. Two neighboring locations are approximately 1m apart. A photo of the meeting room is shown in Fig.~\ref{fig:401K}.

\begin{figure}[htp!]
    \centering
    \includegraphics[width=8cm]{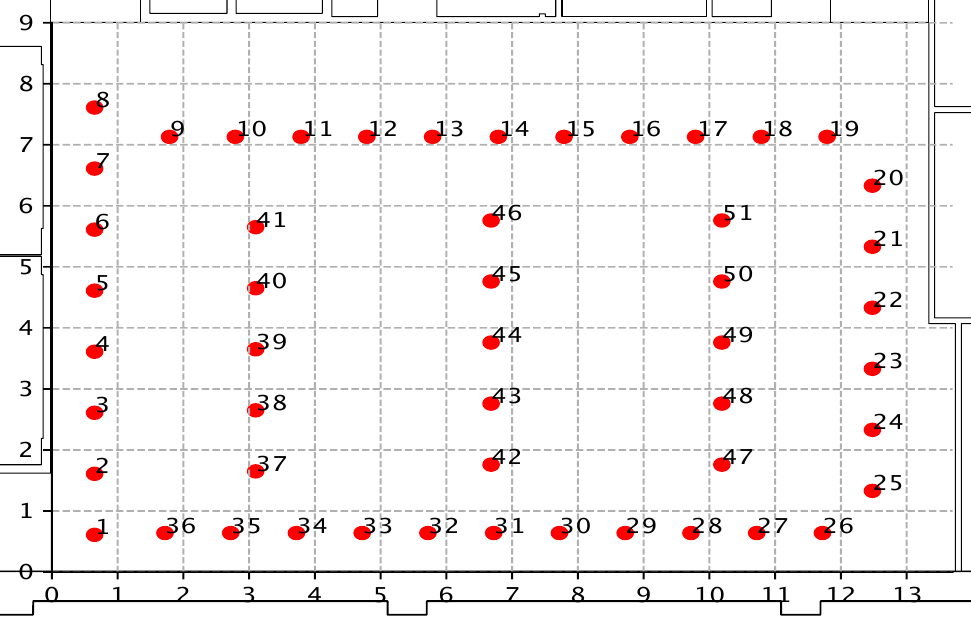}
    \includegraphics[width=8cm]{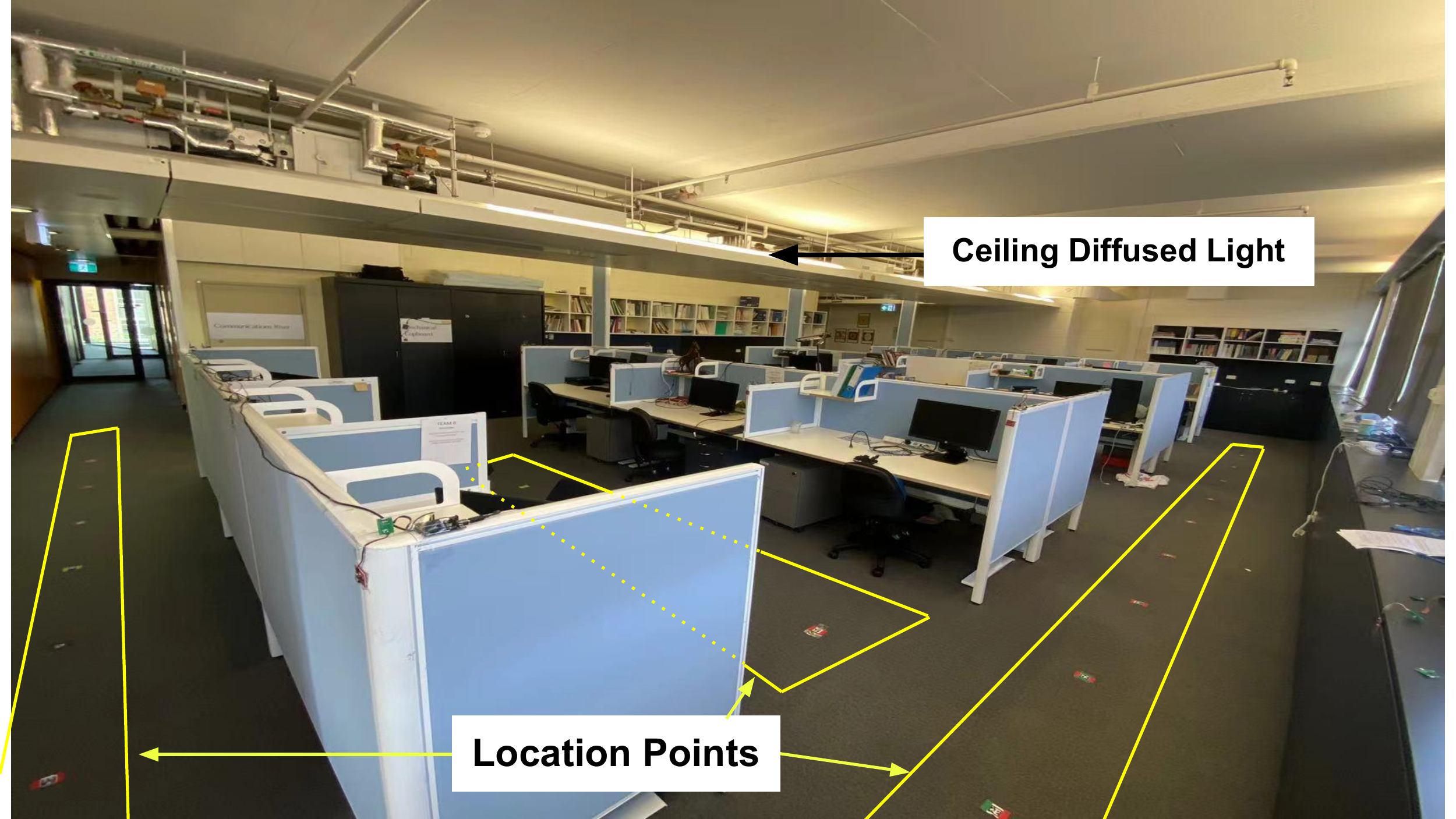}
    \caption{The floor map (top), and the environment of the office evaluation area (bottom). The red points mark the locations, where we collect the lighting condition fingerprints. The distance between two neighboring locations is approximately one meter.}
    \label{fig:412}
\end{figure}

The \textbf{Office}
is a typical open office area with multiple rows of desks and cubicles for the workers. The office has a 13.28m $\times$ 8.33m rectangle shape, but in contrast to the meeting room, it is a much more complex indoor environment. There are a variety of furniture including desks, light-blue partition walls, white plaster walls, metal cabinets, wooden bookshelves, and gray carpets. As highlighted earlier in Fig.~\ref{fig:teaser}, all these different materials have different reflectance profile, which renders the office area more diverse in terms of light spectral distribution. The entire office is illuminated by three rows of Crompton T8 36W 840 fluorescent tube lights plus three circular DULUX D 18 W/840 tube on the ceiling to produce diffused ambient lights to the space. %
Apart from the light switches, we used two floor/desk lamps to realize further variations in lighting conditions. We marked a total of 51 locations spreading around the corridors and between the rows of desks. A photo of the office space along with a location layout is shown in Fig.~\ref{fig:412}.

\begin{figure}[htp!]
    \centering    
    \includegraphics[width=\columnwidth]{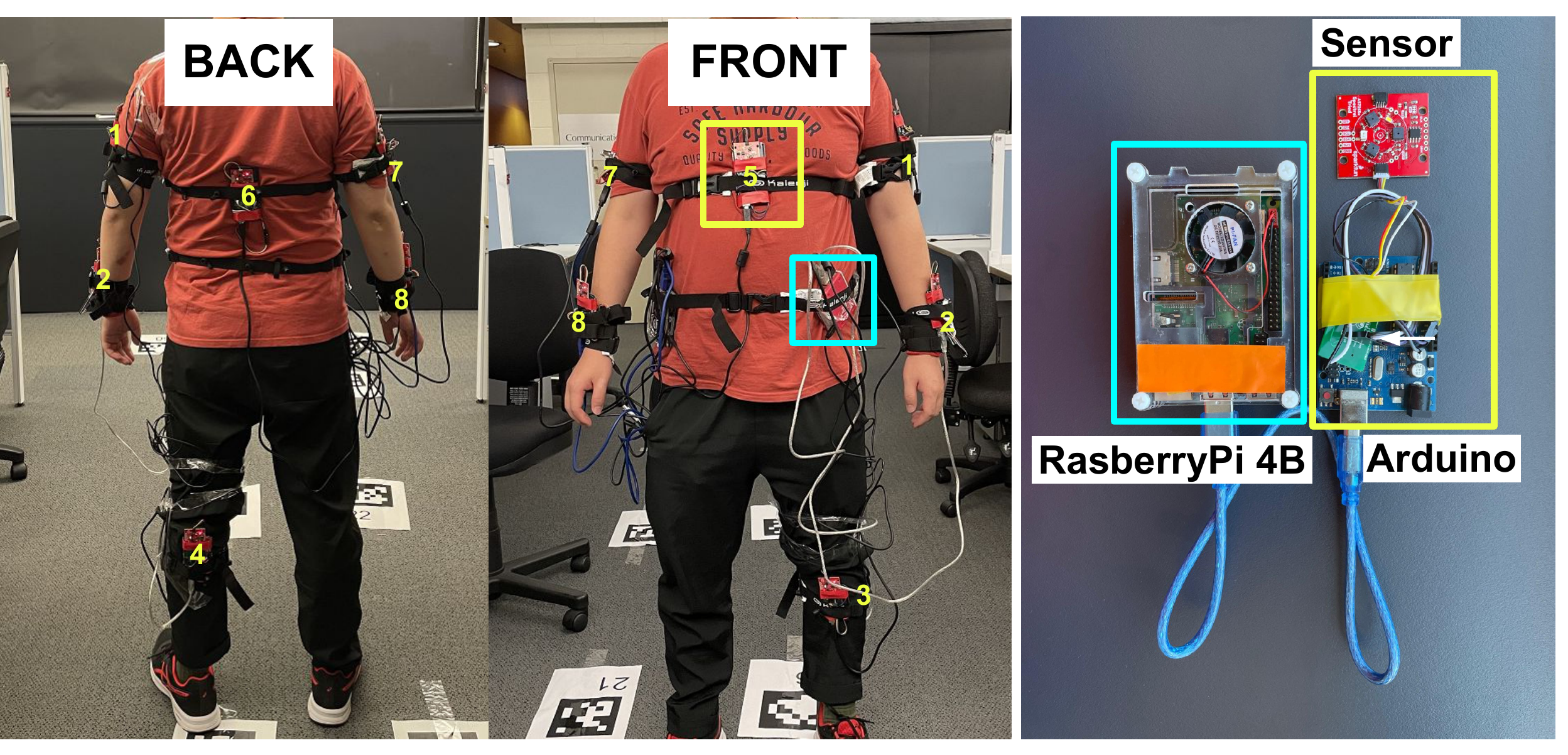}
    \caption{The red points shows the sensors' positions on body, the right picture present how we deploy the sensors.}
    \label{fig:sensor_pos_real}
    \vspace{-0.3cm}
\end{figure}

\textbf{Sensor placement.}
Due to occlusion, the spectral sensor placement location may
affect the localization performance. Furthermore, a better
localization performance may be realized by placing 
multiple spectral sensors in different body locations,
such as chest, back, left arm, left wrist, right arm, right wrist, front of the left calf, and back of the left calf, to capture the 3D spectral fingerprint of
a location. Therefore, we deployed eight spectral sensors in different body
locations shown in Fig.~\ref{fig:sensor_pos_real} and investigate 
their impacts on the localization performance in Sections~\ref{s:number-sensor} and \ref{s:pos-sensor}. 

\textbf{Data collection process.}
For each data collection session in a room (i.e., either the meeting room or the office), a person with eight sensors stands on each marked location for approximately 30 seconds before moving to the next location. 
Upon arriving at a given location, the volunteer first takes a picture of the barcode on the April Tag, which later helps to assign the collected samples to the right locations. There is no interference from background people in the meeting room, but we observed some occasional presence and movement of people in the open office area. The desks in the office area were mostly uninhabited though during the data collection.

Data collection was repeated over multiple days 
to test the algorithm's robustness over time. Table~\ref{tab:day_summary} shows the
detailed experiment setting information such as the number of days, different indoor areas (i.e., office vs meeting room), different lighting conditions and the total amount of samples in each
experiment session.

We implemented the localization neutral network model discussed in Section~\ref{sec:spectral-loc} in Pytorch~\cite{paszke2019pytorch}.
\begin{table}[htp!]
    \centering
        \caption{Data collection statistics for different indoor spaces and lighting conditions.}
    \resizebox{\linewidth}{!}{
    \begin{tabular}{|c|c|c|c|} 
\hline
\begin{tabular}[c]{@{}c@{}}\\Space\end{tabular} & Lighting Condition                                             & Days & Samples  \\ 
\hline
\multirow{3}{*}{Office Area}                    & Default: turn on all lights                                    & 5    & 9,698    \\ 
\cline{2-4}
                                                & Add two floor lamps                                            & 4    & 7,754    \\ 
\cline{2-4}
                                                & Turn off the middle row of ceiling lights                      & 1    & 1,840    \\ 
\hline
\multirow{2}{*}{Meeting Room}                   & Default: Turn on all lights in the room and~adjacent open area & 5    & 1,411    \\ 
\cline{2-4}
                                                & Add one floor lamp                                             & 2    & 739      \\
\hline
\end{tabular}}
\vspace{-0.5cm}
    \label{tab:day_summary}
\end{table}

\subsection{Results}

\subsubsection{Overall localization performance.}
\begin{figure}[htp!]
    \centering
    \includegraphics[width=\columnwidth]{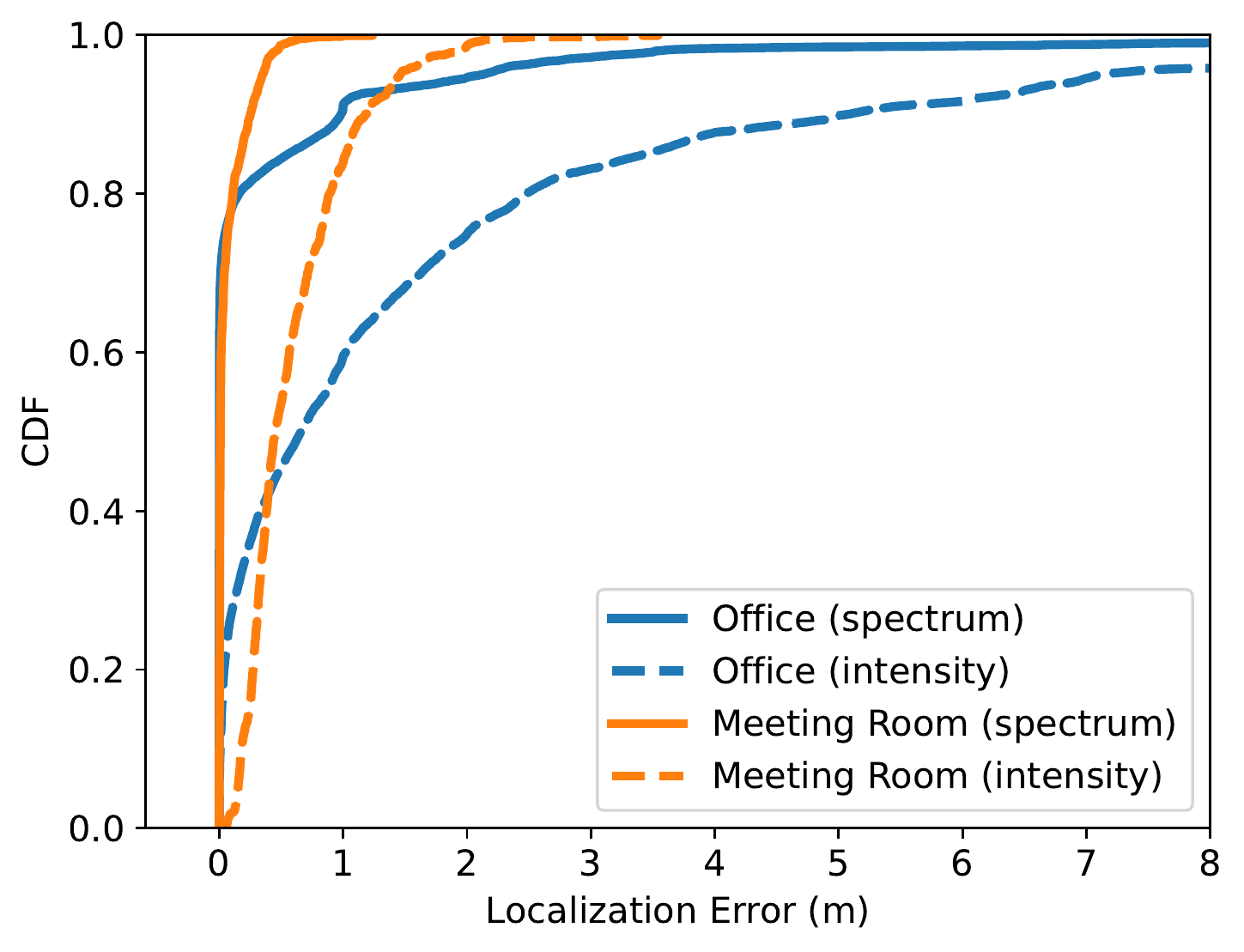}
    \caption{Spectral vs. intensity under default lighting and full sensor view.}
    \label{fig:diffdays}
\end{figure}

We start with benchmarking spectral information against intensity with eight sensors
shown in Fig.~\ref{fig:sensor_pos_real}. Here, we collected data in five different days
 under the default lighting condition in both the office and meeting room (see Table~\ref{tab:day_summary} for the details). Then, we applied the leave-one-day-out testing, i.e., all data from 4 days are used for training, while the data from the remaining day is used for testing. We repeated this five times by selecting a different test day each time and reporting the average localization accuracy in Fig.~\ref{fig:diffdays} with their specific percentile values reported in Table~\ref{tab:percent1}. 
Our results show that, while using light intensity only can achieve a good localization accuracy for the small meeting room, i.e, sub-meter accuracy for both median error and $75^{\rm th}$ percentile, it does not perform well in the large complex environment of the office space even with eight sensors worn on different parts of the body. For example, the $90^{\rm th}$ percentile error is over 5 meters. In contrast, \textit{Spectral-Loc} achieves sub-meter accuracy at the $90^{\rm th}$ percentile in both indoor environments. This demonstrates that \textit{Spectral-Loc} is robust against size and complexity of indoor environments because it exploits environmental (color reflectivity) complexity to produce unique spectral fingerprints of individual locations.
 
\begin{table}[htp!]
    \centering
    \caption{Median, 75th and 90th percentile errors for two rooms under default light conditions.}
    \resizebox{\linewidth}{!}{
    \begin{tabular}{|c|cc|cc|cc|}
\hline
\multirow{2}{*}{\begin{tabular}[c]{@{}c@{}}Room\\ Name\end{tabular}} & \multicolumn{2}{c|}{Median Error (m)}     & \multicolumn{2}{c|}{$75^{th}$\%ile Error(m)}   & \multicolumn{2}{c|}{$90^{th}$\%ile Error(m)}   \\
                                                                     & \multicolumn{1}{c|}{Spectrum} & Intensity & \multicolumn{1}{c|}{Spectrum} & Intensity & \multicolumn{1}{c|}{Spectrum} & Intensity \\ \hline
Office                                                               & \multicolumn{1}{c|}{0}        & 0.66      & \multicolumn{1}{c|}{0.05}     & 2.00      & \multicolumn{1}{c|}{0.98}     & 5.07      \\ \hline
Meeting room                                                         & \multicolumn{1}{c|}{0.01}     & 0.45      & \multicolumn{1}{c|}{0.07}     & 0.82      & \multicolumn{1}{c|}{0.25}     & 1.18      \\ \hline
\end{tabular}}
    
    \label{tab:percent1}
\end{table}

\subsubsection{The impact of the numbers of sensors.}
\label{s:number-sensor}
\begin{figure}
    \centering
    \includegraphics[width=0.8\columnwidth]{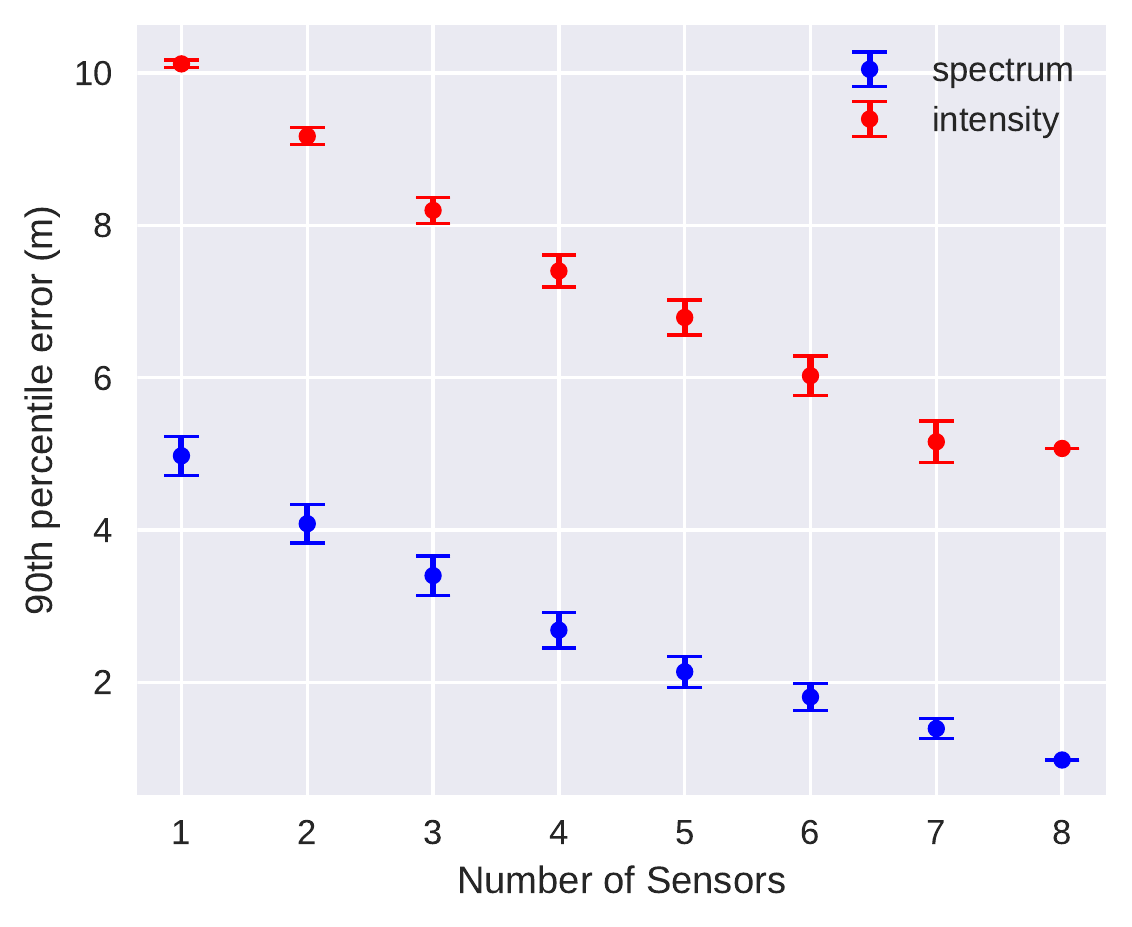}
    \caption{Impact of the number of sensors on localization error (eror bars represent 95\% confidence).}
    \label{fig:diff_num}
    \vspace{-0.5cm}
\end{figure}

To investigate the impact of the number of body-worn sensors to localization performance, we
maintain other factors such as the lighting conditions and the number of wavelength sub-bands
as constants, but vary the number of sensors. Specifically,  
we select a subset of $N,  1 \leq N  \leq 8 $ sensors from eight sensors and use all samples of the selected sensors in the training set to train the neural network. Since we have multiple possible combinations for each $N < 8$, we plot the averages along with the 95\% confidence level for each $N < 8$ in Fig.~\ref{fig:diff_num} for $90^{\rm th}$ percentile error. It shows that the localization error decreases with the increase of the number of sensors for both light intensity and spectrum-based localization algorithms as expected. However, the spectrum-based localization algorithm outperforms its light intensity-based counterpart in each value of $N$ significantly. For example, when the number of sensors is 6 (i.e., $N=6$), the $90^{\rm th}$ percentile error of light intensity-based localization is 6 meters, while that of spectrum-based localization algorithm is less than 2 meters, which represents a threefold improvement. Looking from another angle, we can see that the intensity-based localization would need 8 sensors to achieve a localization accuracy of approximately 5 meters, which can be achieved using only 1 sensor if spectral information was available.

\begin{table}[htp!]
    \centering
          
    \caption{Impact of sensor position on the localization performance (Office).}
        \resizebox{\linewidth}{!}{
\begin{tabular}{|c|c|c|c|} 
\hline
Sensor Position & Median Error (m) & $75^{th}$\%ile error (m)   & $90^{th}$\%ile error (m)  \\ 
\hline
back            & 0.61   & 1.30 & 2.56                 \\ 
\hline
arm-right       & 0.96   & 2.09 & 3.98                 \\ 
\hline
arm-left        & 1.46   & 2.83 & 4.35                 \\ 
\hline
chest           & 0.95   & 2.59 & 4.64                 \\ 
\hline
leg-front       & 1.50   & 3.12 & 5.34                 \\ 
\hline
wrist-right     & 2.54   & 4.30 & 6.10                 \\ 
\hline
wrist-left      & 2.45   & 4.39 & 6.21                 \\ 
\hline
leg-back        & 3.01   & 5.11 & 7.33                 \\
\hline
\end{tabular}}

    \label{tab:single-sensor}
\end{table}

\subsubsection{The impact of sensor positions on the body.}
\label{s:pos-sensor}
Fig.~\ref{fig:diff_num} reports the average localization performance over different combinations for a given number of sensors. Further investigations revealed that the combination of sensors matters. For example, Table \ref{tab:single-sensor} shows that when using a single sensor, $90^{\rm th}$ percentile error can vary widely depending on which sensor is selected. For example, the $90^{\rm th}$ percentile error can be reduced from 7.33 meter (leg-back) to only 2.56 meter by moving the sensor from the leg-back position to the back. This could be perhaps due to the higher height of the back sensor compared to the leg sensor. Indeed, a detailed analysis of all combinations for different values of $N$ revealed that combinations that include the sensors from the upper body generally produce higher localization accuracy compared to the cases when only lower body sensors are used.   
\begin{figure}[htp!]
    \centering
    \begin{subfigure}[b]{0.49\linewidth}
    \centering
    \includegraphics[width=\linewidth]{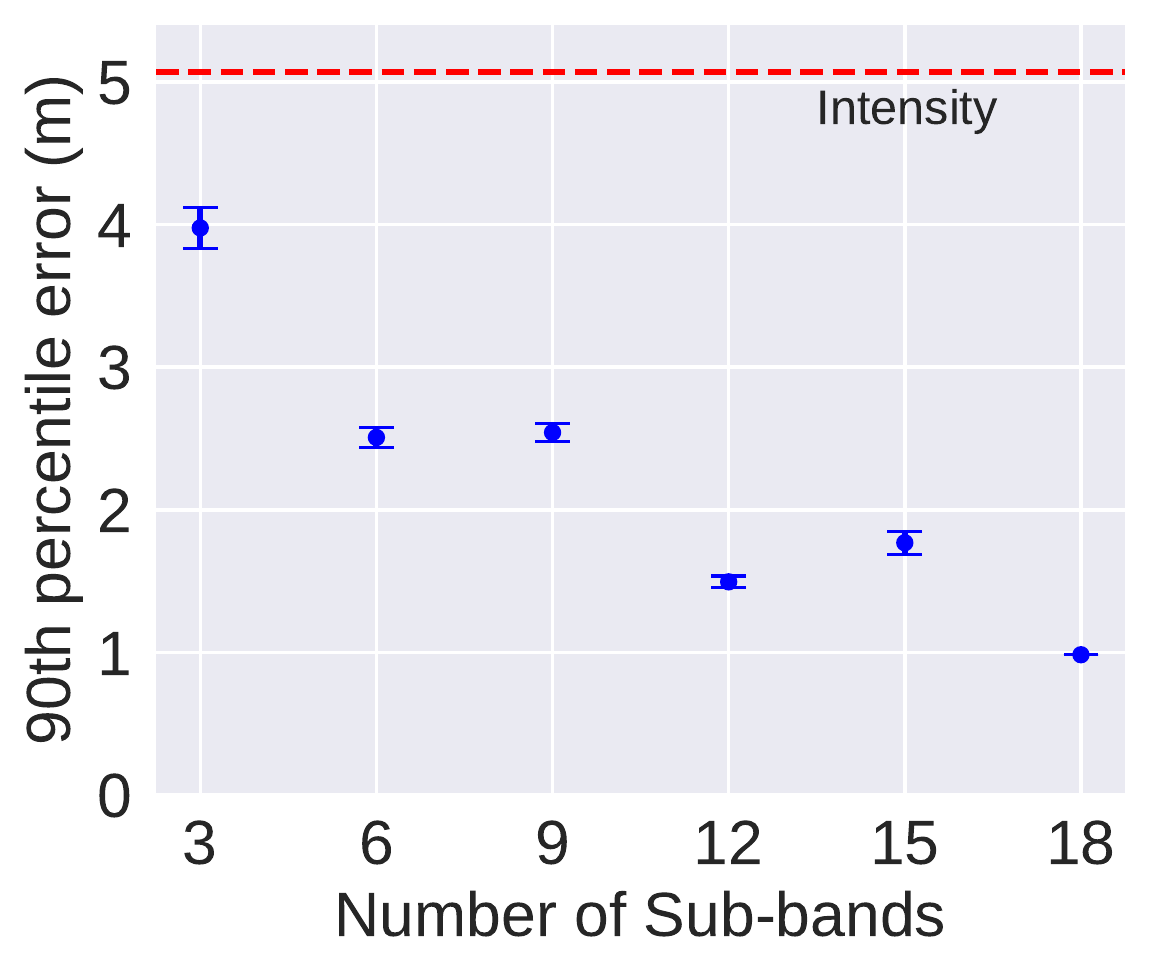}
    \caption{With RGB Restriction}
    \end{subfigure}
    \begin{subfigure}[b]{0.49\linewidth}
    \centering
    \includegraphics[width=\linewidth]{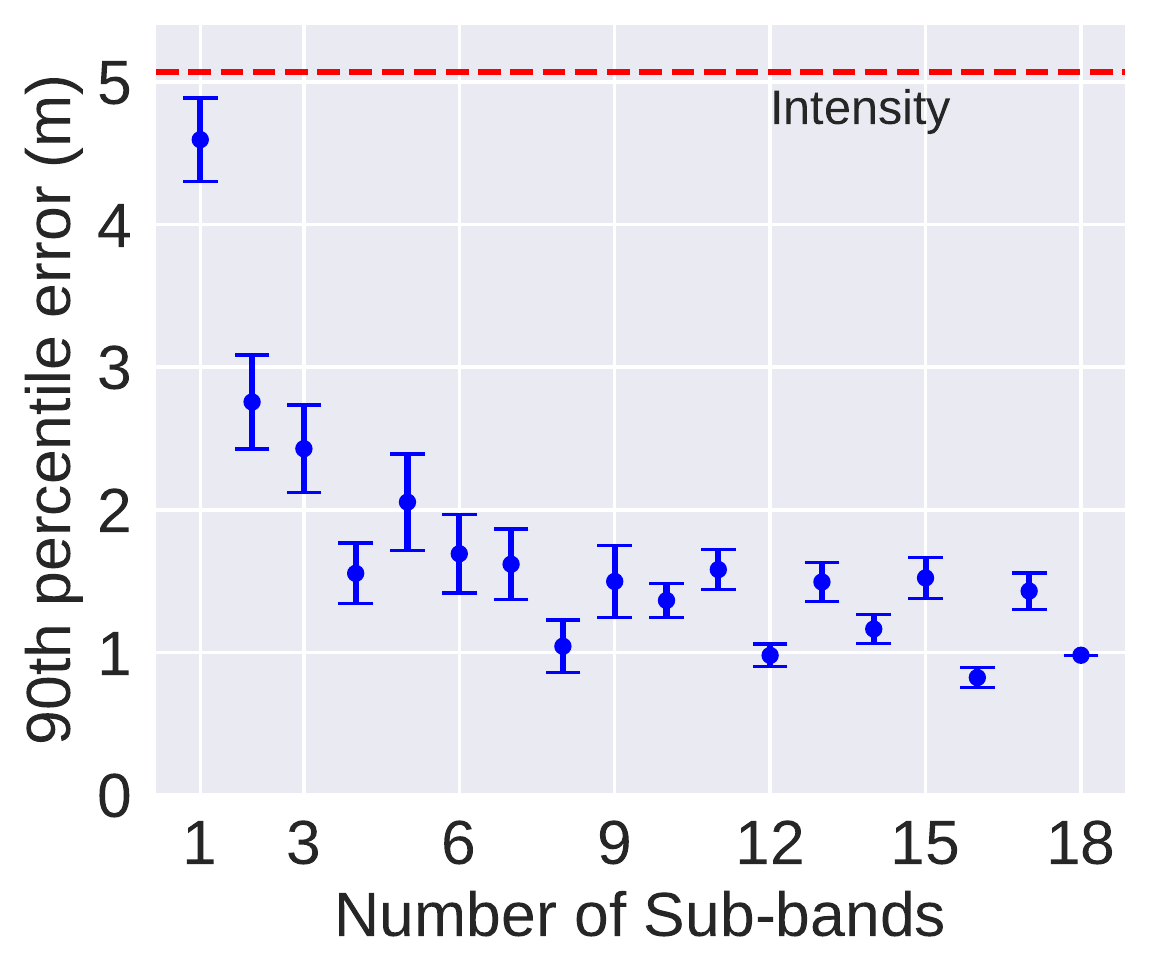}
    \caption{Without RGB Restriction}
    \end{subfigure}
    \caption{Impact of the number of sensing wavelength sub-bands.}
    \label{fig:diff_channel}
    \vspace{-0.5cm}
\end{figure}
\subsubsection{The impact of the number of wavelength sub-bands.}
While AS7265x monitors 18 sub-bands, we wanted to investigate the required number of sub-bands that would produce good indoor localization performance. We analyzed two different policies to select the sub-bands. The first policy, which we call RGB-Restricted, always picks the same number of sub-bands from each of the three primary colors. For example, pick 1 sub-band from RED, 1 from Green and 1 from Blue sub-sensing modules, which gives a total number of 3 sub-bands. Similarly, if we pick 2 from each primary color, then we have a total of 6 sub-bands and so on. 

The second policy has no RGB restriction, i.e., it may have sub-bands selected from anywhere within the visible light spectrum. This policy allows us to select any number of sub-bands in total within the range of 1 to 18. However, unlike the RGB-Restricted policy, this policy may not cover all the primary colours.

For the RGB-Restricted, we evaluate all possible combinations and plot the average along with the 95\% confidence level of the 90-percentile localization error in Fig.~\ref{fig:diff_channel}(a). For the second policy, the total number of combinations is too large, so we randomly pick up $n$ sub-bands, where $1 \le n < 18$, from the 18 available sub-bands 250 times each and plot the average and 95\% confidence bars of the 90-percentile localization error in Fig.~\ref{fig:diff_channel}(b). For $n=18$, we just have one value, so no confidence level is plotted there. We observe that the second policy without RGB restriction performs much better than the RGB-Restricted policy. For this policy, {\bf the localization error decreases non-linearly as we add more channels but the error plateaus with only about 8 sub-bands}; adding more sub-bands beyond 8 does not improve localization any further. 

The analysis in Fig.~\ref{fig:diff_channel}(b) also shows that {\bf even with basic sensors monitoring only a single wavelength, spectral-based localization outperforms intensity-based localization} at the $90^{\rm th}$-percentile --- intensity-based localization error is 5.07m, while spectrum-based localization error is 4.5m with 95\% confidence. This interesting observation can be explained as follows: when one wavelength is decreasing, another may be increasing at the same time, which would keep the total light intensity unchanged. Thus light intensity measurement may miss many subtle differences in the lighting environment, which could be otherwise picked up by an individual wavelength monitor.

\vspace{-0.3cm}
\subsubsection{The impact of lighting interference.}
\begin{figure}[htp!]
    \centering
    \begin{subfigure}[b]{0.49\linewidth}
    \centering
    \includegraphics[width=\linewidth]{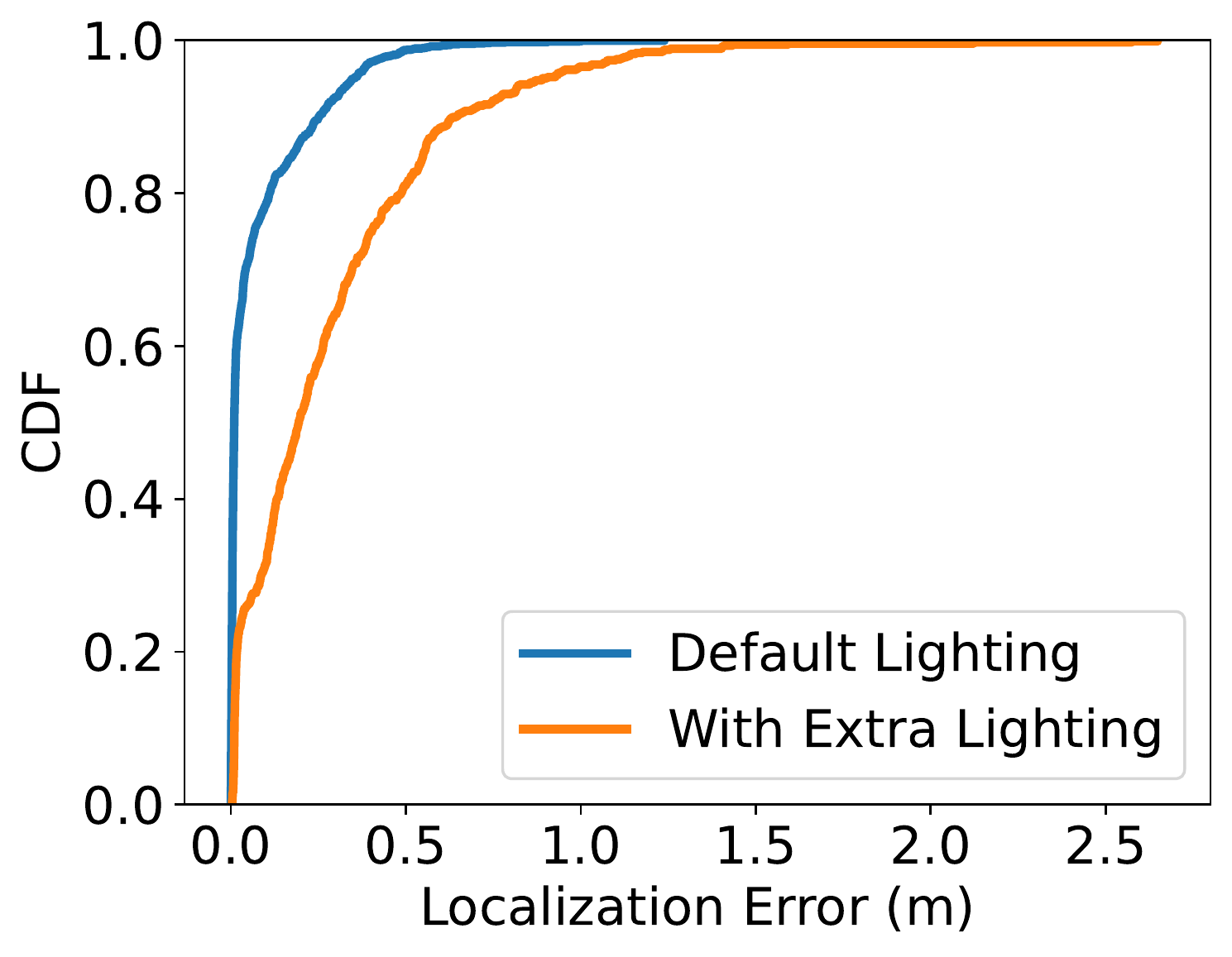}
    \caption{Meeting Room}
    \end{subfigure}
    \begin{subfigure}[b]{0.49\linewidth}
    \centering
    \includegraphics[width=\linewidth]{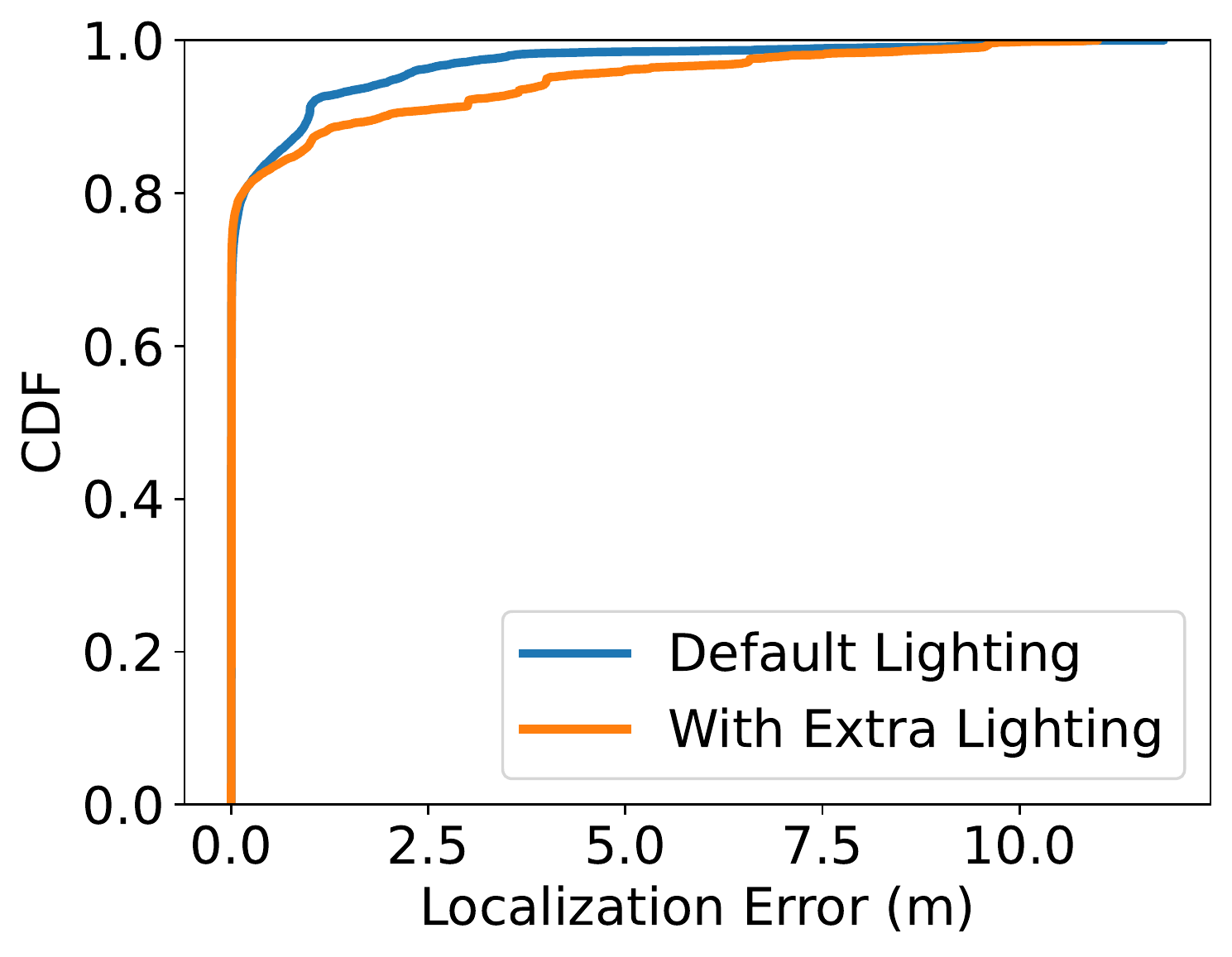}
    \caption{Office}
    \end{subfigure}
    \caption{Impact of lighting interference.}
    \label{fig:interference}
    \vspace{-0.3cm}
\end{figure}

In many indoor events, extra lighting may be added for short-term to improve illumination in specific locations. For example, an attendee in a meeting room may turn on a floor lamp for improved visibility of an exhibit, while an office worker may turn on a desk lamp when trying to focus on a drawing. These extra temporary lighting can be a source of interference when performing localization based on machine learning models that were trained with the default lighting condition. To investigate the performance of \textit{Spectral-Loc} with such lighting interference, we added additional floor/desk lamps in the office and meeting rooms. Specifically, we placed a floor lamp in one corner of the meeting room and collected data in two different days by changing the colour temperature between 2,000k and 4,500k between these two days for increasing the diversity of the lighting interference. Similarly, we placed two lamps at two diagonally opposite desks in the office area and collected data for four days with 2 days in 2,000k and 2 days in 4,500k modes. Then, we trained our localization network using the default non-interfering data and tested with data collected after adding the external lamps. The results are shown in Fig.~\ref{fig:interference}. 

From Fig.~\ref{fig:interference}, we can see that the impact of interference was more pronounced in the meeting room compared to the office area.  This can be explained by the fact that the meeting room is too small and has significantly fewer obstacles compared to the office area, which means the interference directly affected all the locations. In contrast, the office area was large and filled with many occlusions, which had only a minor overall interference to localization. It is important to note, that even with the significant interference in the meeting room, \textit{Spectral-Loc} was able to maintain sub-meter localization accuracy at $75^{\rm th}$ percentile.

\vspace{-0.2cm}
\subsubsection{The impact of lighting failure.}
\label{s:fail}
\begin{figure}[htp!]
    \centering
    \includegraphics[width=0.8\columnwidth]{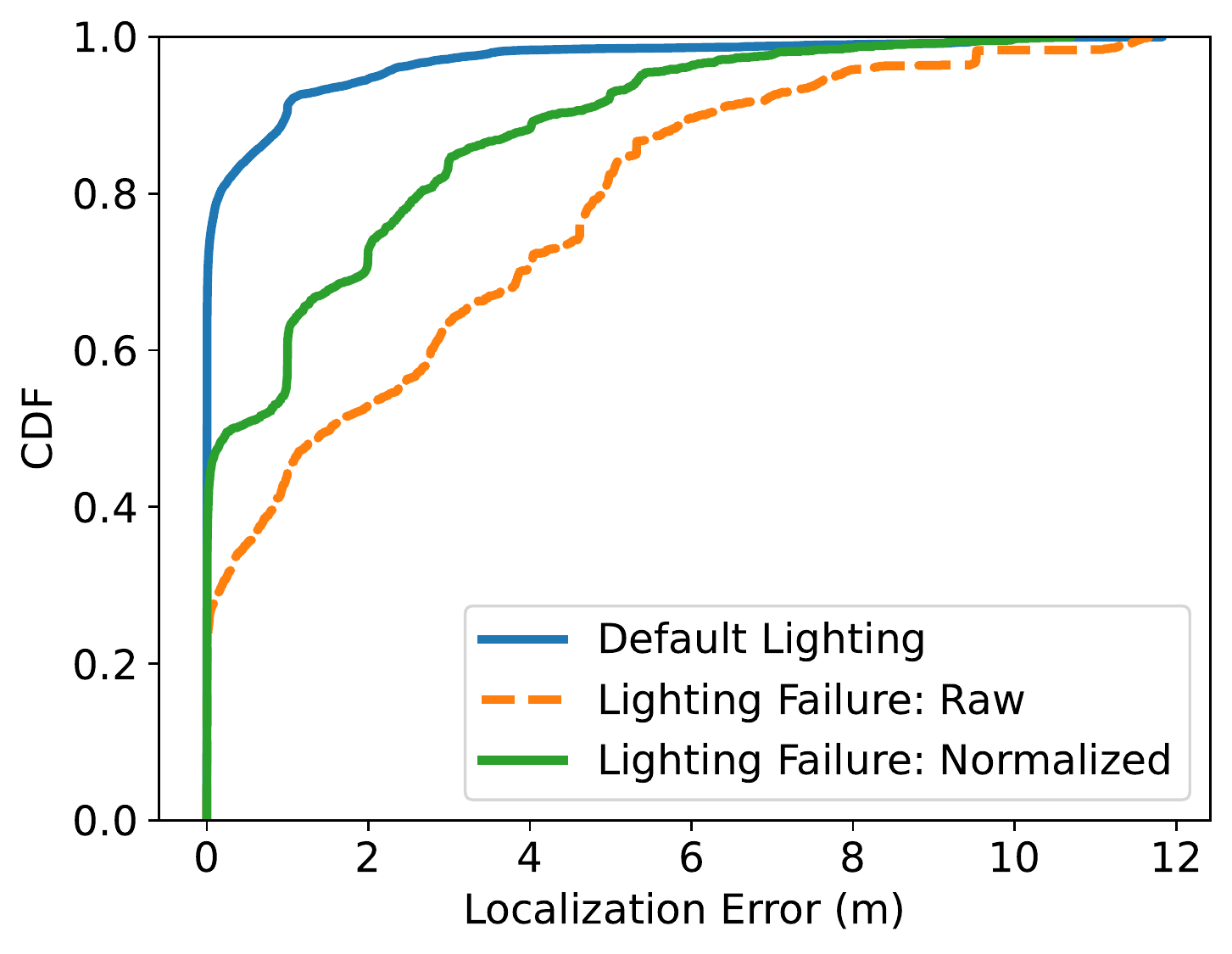}
    \caption{Impact of office lighting failure on \textit{Spectral-Loc} with raw vs. normalized spectral data.}
    \label{fig:stress_test}
    \vspace{-0.3cm}
\end{figure}

However, lights in indoor environments can fail, which requires replacements. Any failed light would reduce the amount of lighting information in the space and hence will have an impact on the localization systems that rely on light information. Thus, until the failed lights are replaced, the localization system would be under stress. We investigated the performance of \textit{Spectral-Loc} under such stress conditions by turning off the middle-row lights in the office area. We then tested the localization system, which was trained with the default lighting condition, with data collected from the darker condition. Results are shown in Fig.~\ref{fig:stress_test}. We observe significant performance drop compared to the default lighting scenario if the machine learning model uses the {\bf raw} spectral sensor data as input. However, by using the proposed {\bf normalized spectral distribution} (see Eqn.~\ref{e:norm}) as the input to machine learning, \textit{Spectral-Loc} can significantly improve its robustness against the lighting failure. Table~\ref{tab:stress} shows that with the normalized spectral features, \textit{Spectral-Loc} can still maintain sub-meter (0.31m) median localization accuracy under the lighting failure, while intensity-based localization has a median error of 3.25m, which is 10$\times$ larger than
that of \textit{Spectral-Loc}.

\begin{table}[htp!]
    \centering
    \caption{Localization performance under lighting failure in the office.}
    \resizebox{\linewidth}{!}{
    \begin{tabular}{|c|c|c|c|c|c|c|} 
\hline
\multirow{2}{*}{\begin{tabular}[c]{@{}c@{}}Lighting \end{tabular}} & \multicolumn{2}{c|}{Median Error} & \multicolumn{2}{c|}{$75^{th}$\%ile Error(m)} & \multicolumn{2}{c|}{$90^{th}$\%ile Error(m)}  \\ 
\cline{2-7}
                                                                                  & Spectral & Intensity              & Spectral & Intensity                         & Spectral & Intensity                          \\ 
\hline
Default                                                                       & 0        & 0.66                   & 0.05     & 2.00                              & 0.98     & 5.07                               \\ 
\hline
Without Middle Row                                                                & 0.31     & 3.25                   & 2.16     & 5.25                              & 4.23     & 6.78                               \\
\hline
\end{tabular}
  }
    
    \label{tab:stress}
    \vspace{-0.5cm}
\end{table}

\section{LIMITATION AND DISCUSSION}
\label{sec:discussion}

In this section, we discuss and reflect on some of the limitations of the current study.

\emph{Impact of human interference.} Humans can block lights and hence interfere with light-based localization. During the data collection in the office area, we observed the occasional movement of people in the area. We can therefore conclude that the presented results are robust against {\bf mild} human interference, which can be expected in meeting rooms, open office areas, residential spaces, and so on. Further study would be required to understand the full impact of human interference in crowded indoor environments, such as shopping malls or underground train stations. 

\emph{Localization for walking users.} In this work, we considered localizing users standing at a particular location, hence using light spectral data gathered from a fixed location for fingerprinting. Recent works with light intensity-based localization~\cite{zhao2017navilight} has demonstrated that localization accuracy can be significantly improved by considering a vector of multiple light intensity values collected from a series of locations in the user's walk trace. We believe that the reported localization accuracy values for \textit{Spectral-Loc} could also be improved further for walking users with such trace data, but such studies would be orthogonal to the current study.

\emph{Daytime localization.} Current study is limited to night-time localization, where we do not experience interference from sunlight. Daytime light-based localization, however, is much more challenging as factors such as weather, climate, the orientation of the building, window sizes and quality, etc. can significantly change the spectral distribution and intensity in the building. Hence, realizing localization robustness to different lighting conditions in daytime remains an open problem for any light-based localization and \textit{Spectral-Loc} is not immune from it.

\emph{Sensor and user orientation.} In our experiment, the positions of sensors on the human body are fixed, and the subject always faces the same direction when collecting data to realize a more controlled setting. However, we did fit the human body with 8 sensors with different orientations. For example, the back sensor is facing 180 degrees from the chest sensor. Therefore, the localization results reported for multiple sensors would be robust against user orientation. 

\emph{Sensed spectrum.} We used AS7265x whose sensing is limited within 410nm-940nm, which we believe was adequate to capture the lighting and the reflections within our experimental indoor spaces. For more diverse indoor spaces comprising a wider range of lights and materials, it may be worth sensing a wider spectrum, such as using AS7341~\cite{as7341} which can sense from 350nm to 1,000nm. 

\emph{Energy consumption.} Spectral sensing would have higher energy consumption compared to the basic intensity sensing due to sampling and signal processing from multiple light
wavelength sub-bands. However, the analysis presented in this paper has shown that the impact of the number of sub-bands on localization accuracy is non-linear, i.e., the localization error can be reduced significantly by using a few sub-bands but adding more sub-bands thereafter provides only marginal improvements. This discovery indicates that the energy consumption of the proposed light spectrum-based localization can be reduced significantly by using sensors that measures a smaller number of sub-bands. For example, AS7265x measures 18 sub-bands and requires a supply voltage of 2.7-3.6V, whereas AS7341 measures 8 sub-bands using only 1.8V.

\section{RELATED WORK}
\balance
\label{sec:related}
Over the past decades, researches focused on localization techniques based on many different types of radios, for instance, WiFi \cite{abbas2019wideep,chen2020fido,chen2020fido}, Bluetooth \cite{zhuang2016smartphone,bianchi2018rssi}, Lo-Ra  \cite{liu2021seirios}, ZigBee \cite{sugano2006indoor,niu2015zil,bianchi2018rssi}, and Ultra-Wide band (UWB) \cite{prorok2014accurate,hanssens2016indoor}.  These wireless-based localization methods can achieve localization accuracy from $cm$ level to $m$ level depending on different systems. However, the wireless signal based approaches are susceptible to interference from other wireless communications in the environment.

As light is readily available and densely deployed in most indoor spaces, there has been a recent interest in utilizing visible light for indoor localization. Light-based localization systems can be broadly categorized as modulated, where lighting infrastructure is modified, and unmodulated, which enables localization using only the existing ambient light.

\textbf{Modulated Light.} 
Luxapose \cite{kuo2014luxapose} modified LED with pulse width modulation.
Spatial beams has been identified with unique timed sequence of light signals in Spinlight \cite{xie2015spinlight}. For light polarization, Celli \cite{wei2017celli} projects interference-free polarized light beams and PIXEL~\cite{yang2015wearables} using liquid crystal to modulate the polarized light. SmartLight \cite{liu2017smartlight} uses the light splitting property of convex lens. EyeLight \cite{nguyen2018eyelight} uses reflection property of light for localizaiton. 
Those different types of techniques have been explored to modulate the light source, which can enhance the distance relationship between the users (receivers) and light sources (transmitters). For example, FogLight \cite{ma2017foglight} used the off-the-shelf digital projector combined with a light sensor. It projected the grey-coded binary pattern using the alternating property of the Digital Light Processing, then decoded and transformed this binary pattern and sent the position via their WiFi module. By leveraging white and black to represent 1 and 0 inside the projection area, the pixel can be represented as a sequence of binary digits. They can achieve a high accuracy result in localization with the light projection, which achieves 0.3 cm for $90^{\rm th}$ percentile distance errors. Apart from using intensity information, \cite{tian2018augmenting} augmented indoor inertial tracking by reusing existing indoor luminaries to project a static light polarization pattern in the space. It uses polarizers and birefringent films to create an imperceptible grid pattern of light polarization. As the polarized light is imperceptible to human eyes, a simple color sensor is used to decode the polarized colorful pattern. However, the spectrum changes are caused by the modulated light source. How to utilize spectral information with unmodulated ambient light has not been investigated yet.

\textbf{Unmodulated Light.} For the unmodulated scenarios, the critical problem is finding location-related features. One way is regarding the light source as land markers and acquiring its discriminative features \cite{hu2015lightitude,zhang2016litell,munir2019passive}. LiTell~\cite{zhang2016litell} captures the unique frequency of fluorescent lights through cameras to match the known location. The other way is extracting information from light. Intensity is one of the primary measurements of light, while it is a scalar value that lacks unique identification of locations. To address this problem, NaviLight \cite{zhao2017navilight} observed intensity changes with a specific walking path. They used light intensity as a fingerprint similar to people using the Received Signal Strength Indicator (RSSI) of WiFi as a fingerprint. However, light intensity is ambiguous over the air. It is not enough to only use it to do the fine-grained localization. Hence, NaviLight firstly uses a k-nearest neighbours classifier to do a coarse-grained localization. After that, it collected IMU data from users' movement, divided the fingerprint vector into small chunks, and then mapped it in the light intensity floor map for fine-grained localization. Combing other sensors can be another way to help identify locations, for instance, using magnetic sensors \cite{wang2018deepml,wang2020indoor}. They used the magnetic and intensity data to create bimodal images. These are the main two ways to extract more location related information. Generally, compared with the modulated light source, the cost of the non-modulated light source method is relatively smaller, but the processing of the received light signal and the algorithm of localization will be more complex and computationally intensive.

With the light-based localization method, the challenge such as interference from blocking between users and light sources, interference from the ambient light sources, not working in light off mode happened in most systems. Some studies try to address these issues. For example, EyeLight \cite{nguyen2018eyelight} addressed the Line of Sight problem by leveraging shadows. However, its price for this is a reduction in localization accuracy. These tasks are still to be addressed by future works.

\textbf{Spectral sensors.} Spectral sensors can be divided into two categories: RGB color sensors and multiple sub-band sensors. The RGB sensors use the interference filters to measure the absolute values of three color sub-bands. After calibration, the sensitivity for this kind of sensor is similar to human eyes. The other type of spectral sensors is multi-bands spectral sensors, such as AS7265x~\cite{spectralsensor} we used, which measure the radiant power for multiple sub-bands. With the help of spectral sensors, we can tell the orange light is mixed of red and yellow light or pure orange light. They have been used in the smartphones, for instance, Xiaomi \cite{xiaomi}, Huawei~\cite{huawei}, where the spectral sensor can acquire accurate white-balancing to improve the rendition of color in photography. Besides, the spectral sensors can be applied in video equipment and virtual reality devices. Through the spectral readings, the light source can be analyzed to differentiate the light source type, for instance, LED, solar, etc. In this way, according to the spectrum, the accurate color compensation for different light sources can be proposed to improve the video rendering effect, which makes the devices robust to different light conditions. Besides image rendering, spectral sensors have been recently used to detect radiation~\cite{leon2021measuring} and airport lighting types for automated lighting maintenance~\cite{polish2021}. To the best of our knowledge, the use of spectral sensors for indoor localization is yet to be reported in the open literature.

\section{CONCLUSION}
\label{sec:conclusion}

We have studied, for the first time, the potential for indoor localization using the light spectrum information extracted from the ambient lights. Our study has confirmed that light spectrum information that can be measured using low-cost and low-power COTS sensors is capable of fingerprinting typical indoor locations much more accurately than that of the light intensity information. We have developed signal processing and machine learning required to exploit light spectral information for indoor localization. In our experiments with a meeting room as well as a large open office space, we were able to realize sub-meter median accuracy for the proposed light-spectral-information-based localization system using only a single sensor on the user body. These encouraging results combined with recent proliferation of spectral sensors in commodity mobile devices are expected to open up new avenues for precise indoor localization using only the existing lighting infrastructure.

\begin{acks}
This work was partially funded by the Australian Research Council Discovery Project DP210100904, UNSW Scientia
PhD Scholarship Scheme and CSIRO Data61 PhD Scholarship Program.
\end{acks}

\bibliographystyle{ACM-Reference-Format}
\bibliography{reference}
\end{document}